\documentclass[article,11pt,leqno]{elsarticle}

\usepackage[utf8]{inputenc} 		% applemac or utf8
\usepackage[T1]{fontenc}
\usepackage{lmodern}
\usepackage{amssymb,amsmath,stmaryrd}
\usepackage{amsthm}
\usepackage{dsfont}
\usepackage{longtable,booktabs}
\usepackage{enumerate}
\usepackage{multicol}
\usepackage{lineno,hyperref}
\hypersetup{colorlinks=true,citecolor=magenta,urlcolor=blue}
\usepackage{epsfig}
\usepackage{pstricks}
\usepackage{graphicx}
\usepackage{algorithmicx}
\usepackage{fullpage}
\usepackage{lscape}
\usepackage{multirow}
\usepackage{subcaption}
\usepackage{algorithm}
\usepackage[noend]{algpseudocode}
\input epsf

\newcommand{\ba}{\begin{eqnarray}}
\newcommand{\ea}{\end{eqnarray}}

\newenvironment{Proofc}[1]{\smallskip\par\noindent\textbf{#1}\quad}%
  {\hfill$\Box$\bigskip\par}

\nonstopmode

\graphicspath{{figures/}}

\journal{Journal of Computational Science}

\usepackage{ulem}

\begin{document}
\begin{frontmatter}


\title{Real-time Mobile Sensor Management Framework for city-scale environmental monitoring}



\author[1]{Kun Qian}
\ead{kunqian@utexas.edu}
\author[1]{Christian Claudel}
\ead{christian.claudel@utexas.edu}
\cortext[cor]{Corresponding author. Tel. : (737)484-2035}

\address[1]{The University of Texas at Austin, Austin, TX 78712, USA.}

\address{Texas, United States}

\begin{abstract}
Environmental disasters such as flash floods are becoming more and more prevalent and carry an increasing burden to human civilization. They are usually unpredictable, fast in development and extend across large geographical areas. The consequences of such disasters can be reduced through better monitoring, for example using mobile sensing platforms that can give timely and accurate information to first responders and the public. Given the extended scale of the areas to monitor, and the time-varying nature of the phenomenon, we need fast algorithms to quickly determine the best sequence of locations to be monitored. This problem is very challenging: the present informative mobile sensor routing algorithms are either short-sighted or computationally demanding when applied to large scale systems. In this paper, a real-time sensor task scheduling algorithm that suits the features and needs of city-scale environmental monitoring tasks is proposed. The algorithm is run in forward search and makes use of the predictions of an associated distributed parameter system, modeling flash flood propagation. It partly inherits the causal relation expressed by a search tree, which describes all possible sequential decisions. The computationally heavy data assimilation steps in the forward search tree are replaced by functions dependent on the covariance matrix between observation sets. Taking flood tracking in an urban area as a concrete example, numerical experiments in this paper indicate that this scheduling algorithm can achieve better results than myopic planning algorithms and other heuristics based sensor placement algorithms. Furthermore, this paper relies on a deep learning-based data-driven model to track the system states, and experiments suggest that popular estimation techniques have very good performance when applied to precise data-driven models. The data and code can be freely downloaded from~\footnote{\url{https://drive.google.com/drive/folders/1gRz4T2KGFXtlnSugarfUL8r355cXb7Ko?usp=sharing}}.
\end{abstract}

\begin{keyword}
Environmental Monitoring \sep Mobile Sensing \sep Data Assimilation \sep Data-driven Model 

% MSC codes here, in the form: \MSC code \sep code
% or \MSC[2008] code \sep code (2000 is the default)

\end{keyword}

\end{frontmatter}

\section{Introduction}
\label{sec:intro}
Floods are one of the most damaging natural disasters, accounting for 31$\%$ of economic losses resulting from all the natural disasters. The ten costliest floods between 1989 and 2014 caused an estimated US $\$$187 billion losses and 13,597 casualties overall \cite{trigg2016credibility}. Several flood mitigation strategies exist, including flood channels, early warning systems \cite{krzhizhanovskaya2011flood} and temporary flood barriers~\cite{few2003flooding}. Among all these strategies, flood monitoring and prediction is one of the most effective choices in terms of cost to benefit ratio. Such an early warning and emergency management system relies on a model that can provide a real-time prediction of flooding, and a sensor network to provide measurements and correct the model results. This article proposes a real-time algorithm to control robotic flood sensors over large areas. This algorithm is not restricted to flood monitoring but can be generalized to different natural hazards monitoring tasks such as wildfire, oil spills and other problems involving large scale distributed parameter systems. We also perform numerical experiments that demonstrate the effectiveness of the proposed method, which achieves near-optimal performance in minimizing estimation errors and gathering system information.

\begin{figure}
    \centering
    \includegraphics[width =\textwidth]{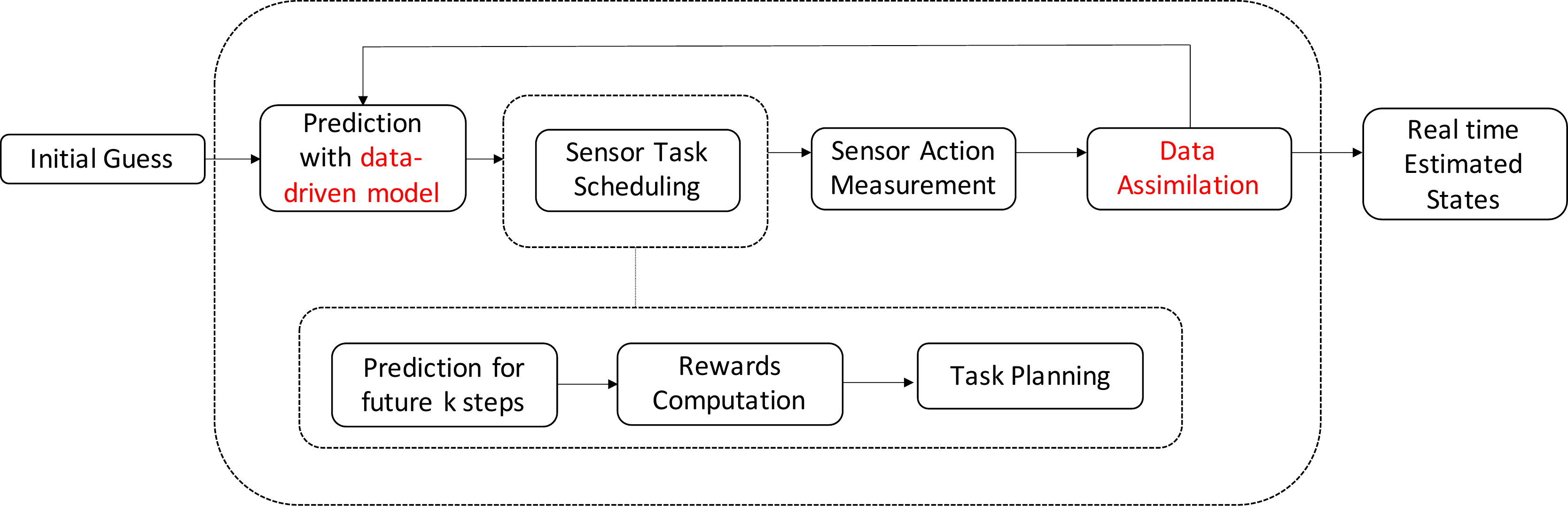}
    \caption{Overview of sensor placement problem investigated in this article. Starting from an initial guess of the environmental states, the deep learning based data-driven model~\cite{qian2019physics} is applied to predict future states. Using these results, the sensor task scheduling module plans the sensing actions (trajectories and sensing schedule) for mobile sensors. Sensors then follow the planned tasks to take measurements at desired locations. An Ensemble Kalman filter is employed to assimilate measurements. The focus of this article is to introduce an informative and nonmyopic sensor task scheduling algorithm which can be implemented in real-time for such large scale systems.}
    \label{fig:generalflow}
\end{figure}

The workflow of the computational framework developed in this article is presented in Figure~\ref{fig:generalflow}. Several difficulties exist when attempting to implement this framework in real-time. The first difficulty involves the Partial Differential Equation (PDE) based prediction model, which is computationally challenging. Furthermore, the sensor task scheduling does not scale well to large problems. The sensor task scheduling is a sequential decision making process in which control problems are coupled with estimation. The most popular methods currently available, such as pruning forward searching tree~\cite{atanasov2014information, vitus2012efficient}, cannot solve the scheduling problem in real-time in large scale problems over some reasonable time horizon.

To enable real-time flood prediction, \cite{qian2019physics} introduced a physics informed deep learning flood prediction model that can generate city-scale high resolution flood predictions in real-time. The present work builds on this prediction model~\cite{qian2019physics}. 
% For the data assimilation part, Ensemble Kalman Filter(EnKF) is proved to be the most practical and useful method for the state estimation of large scale nonlinear system~\cite{evensen1994sequential, houtekamer2001sequential, hamill2001distance, houtekamer1998data}.

% The new algorithm employs the full-power of model prediction. It reduces the computation amount through skipping the most time consuming part-data assimilation in sensor tasks scheduling process. However, in order to make up for the missed information by skipping the data-assimilation steps, the new algorithm added an empirical function to consider the causal relation between actions at different steps.
Considerable attention has been given to the problem of designing fixed sensor networks for environmental monitoring~\cite{kerkez2012design, weimer2008relaxation}. Based on the submodular property of the mutual information function, \cite{krause2008near} derived a greedy algorithm to select sensor locations for fixed sensor networks. \cite{joshi2008sensor} transformed the sensor location selection problem into a convex optimization problem. 

In contrast to fixed sensor networks, mobile sensors have many advantages~\cite{tokekar2013tracking, tokekar2016sensor, choi2009adaptive}. Since natural hazards usually happen over large spatial scales and tend to be unpredictable, it would be expensive and inefficient to maintain a large number of fixed sensors. However, one of the difficulties associated with mobile sensors is the problem of sensor placement and scheduling. Many heuristic-based task scheduling algorithms~\cite{neumann2011artificial, zhang2007adaptive} have been developed in the past, such as following Voronoi graphs to cover an area~\cite{schwager2015robust} or reformatting the problem into traveling salesman-like problems~\cite{tokekar2016sensor, yu2014correlated}. Beyond these heuristics, \cite{singh2007efficient} proposes a myopic informative path planning algorithm for the robotics to explore a spatial distribution. \cite{singh2009nonmyopic} extends the work~\cite{singh2007efficient} to a non-myopic informative path planning algorithm. However, a limitation of the aforementioned works is that they assume that the events follow a Gaussian Distribution in space. In contrast with state-space models, Gaussian Processes are not accurate enough to model complex, highly nonlinear systems such as flash floods, though the computational requirements are much higher when dealing with state-space models. It has been shown in~\cite{vitus2012efficient} that for linear state-space models, one can prune branches for the search tree and significantly reduce computational time. \cite{schlotfeldt2018anytime, atanasov2014information, jawaid2015submodularity, asghar2017complete} followed the idea in~\cite{vitus2012efficient} to develop practical algorithms for environmental monitoring and searching applications. However, these algorithms are implemented in real-time at the cost of falling back to myopic scheduling strategies, and myopic strategy is not desired in city-scale environmental monitoring which would be shown in Section~\ref{sec:experiments}. 

% This paper would integrate dynamical system of the environment and path planning of mobile sensors together to work out a real-time non-myopic informative mobile sensor task scheduling algorithm. 

% There are recent works considering dynamically learning the parameters of Gaussian Process during sensing process~\cite{luo2018adaptive, ma2017informative}. \cite{le2009trajectory} introduces a search tree style forward value iteration algorithm to plan path for mobile sensors with Gaussian Process Model.

Another related field to this article is sensor management for object tracking. In sensor management problems, the objective is usually to maximize the information gain~\cite{hero2007foundations, hero2011sensor} with finite sensing resources. \cite{dames2015autonomous, charrow2014approximate, hoffmann2009mobile} focuses on target monitoring with approximate mutual information. \cite{lauri2014stochastic} formulates the active sensing problem into a Partially Observed Markov Decision Process(POMDP). \cite{choudhury2017learning} introduces a novel data-driven approach via imitation learning. The methods used in object tracking do not apply well to distributed parameters systems, however. Often, the dimensionality of the system to be modeled is usually much larger in distributed parameters systems.

This article proposes an informative and non-myopic mobile sensor task scheduling algorithm that is computationally efficient and can achieve near-optimal performance over simulated experiments. For this, we use predictions based on a deep learning model derived from conventional PDE flood propagation models. The task scheduling algorithm is based on a simplified forward search tree. It reorganizes the previously exponential growing trees and skips the data assimilation steps in the tree. An empirical function depending on the correlation of different observation sets is added to compensate for the skipped data assimilation steps in the planning process. Although the number of action sequences to be checked would still grow exponentially with the prediction horizon, this algorithm saves most of the data assimilation steps which are slowest to compute. Numerical experiments are conducted for the entire state estimation framework with a special focus on the task scheduling algorithm. It is shown that this sensor task scheduling algorithm achieves near-optimal performance. 
% \textcolor{red}{This paper applies the algorithm to city-scale flood estimation which has a high resolution of 30000 dimensions. It's considered to be a large scale system compared with tracking tasks. Compared with data assimilation tasks for ocean studies which involve millions of states, the dimension of case studied here is of relative small dimension.}

% We reorganize the original search tree which has a exponential growing branches to an approximated search tree which has branch numbers linear to prediction horizon. Some heuristics based functions are added to make up the missed information caused by approximation. The numerical experiments indicate that this algorithm provides near optimal performance in tracking an environmental event. 

Section~\ref{sec:ltrreview} presents in detail the problem that the article intends to solve. In Section~\ref{sec:methods}, each part in the mobile sensor framework is explained with a focus on the real-time mobile sensor task scheduling algorithm. Numerical experiment results are shown in Section~\ref{sec:experiments}. Section~\ref{sec:conclusion} presents additional insights and perspectives for future work.
\section{Problem Description}
\label{sec:ltrreview}
We follow and extend the problem formulation of active information acquisition problem introduced in \cite{atanasov2014information}. The formulation is adapted to better suit nonlinear high dimensional distributed parameters systems. 

Consider mobile sensors whose dynamics are governed by 
\begin{equation}
\label{equ:reachableconstraint}
    \omega(t+1) \in \mathcal{A}(\omega(t))
\end{equation}
in which $\omega(t)$ is the states of mobile sensors at time $t$ and $\mathcal{A}(\omega(t))$ denotes the reachable set from $\omega(t)$.

The goal for mobile sensors is to monitor the states of a distributed parameters system. Suppose that the estimated states of the system is denoted as $b_S(t|Z_{1:t-1})$ in which $Z_{1:t-1}$ means the observations from time $1$ to $t-1$, the dynamics of the distributed parameters system is written as 
\begin{equation}
\label{equ:dynamicsconstraint}
    b_{S}(t|Z_{1:t_0}, \hat{Z}_{\omega(t_0)},\dots, \hat{Z}_{\omega(t-1)}) = f(b_{S}(t-1|Z_{1:t_0}, \hat{Z}_{\omega(t_0+1)},\dots, \hat{Z}_{\omega(t-1)}), u(t))
\end{equation}
in which $u(t)$ is an uncontrollable input term to the distributed parameters system and $\hat{Z}$ is used to denote the expected observation value instead of a real observation value. 

We suppose that the observation model is linear, such that it can be described as 
\begin{equation}
\label{equ:obs}
    \hat{Z}_{\omega(t)} = H_{\omega(t)}(b_{S}(t|Z_{1:t_0-1}, \hat{Z}_{\omega(t_0)},\dots, \hat{Z}_{\omega(t-1)}))
\end{equation}
in which $\hat{Z}$ denotes the observation value extracted from belief states, and $H_{\omega(t)}$ denotes the observation model under the sensor state $\omega(t)$. 

The major difference with previous formulation~\cite{atanasov2014information} is that previous formulation assumes that the system to be tracked is linear, however, we consider a nonlinear system. The differences arises as linear systems only need to consider the evolve of covariance matrix of estimated system while nonlinear systems need to consider both the mean value and covariance matrix because the dynamics changes with different mean values. Thus, we use a generalized expression for the measurement update.

\begin{equation}
\label{equ:mupdate}
b_{S}(t|Z_{1:t_0}, \hat{Z}_{\omega(t_0)},\dots, \hat{Z}_{\omega(t-1)}, \hat{Z}_{\omega(t)}) = M(b_{S}(t|Z_{1:t_0}, \hat{Z}_{\omega(t_0)},\dots, \hat{Z}_{\omega(t-1)}), \hat{Z}_{\omega(t)})
\end{equation}

in which $M$ denotes the function of data assimilation. In this paper, we assume that the distribution of states of the distributed parameters system follows Gaussian distribution and Ensemble Kalman Filter will be applied for the data assimilation step. However, the function $M$ can be any data assimilation methods.

Given all the constraints, we give the objective function.

\begin{equation}
\label{equ:MSEerror}
\min_{\omega(t_0 + 1), \dots, \omega(t_0 + T)}\frac{1}{T} \sum_{t=t_{0}}^{T + t_{0}} \gamma^{t - t_0} \mathbf{R}(\omega(t), b_{S}(t|Z_{1:t_0-1}, \hat{Z}_{\omega(t_{0})}, \dots, \hat{Z}_{\omega(t)}))
\end{equation}

in which $\gamma$ is a discounting coefficient and $\mathbf{R}$ denotes any reward function which can measure the uncertainty of the system. Note that $\mathbf{R}$ can also be the information gain of each sensing action. However, the $\min$ should be changed to $\max$ if information gain is used as the metric. The expression of objective function is also more generic than~\cite{atanasov2014information}. The detailed information of function $\mathbf{R}$ is discussed in Section~\ref{sec:methods}.

\section{Methodology}
\label{sec:methods}
% Technique details are introduced for each part of the real-time mobile sensing frame~\ref{fig:generalflow} in this Section. The emphasis is to introduce the real-time mobile sensor tasks scheduling algorithm. This frame is based on the deep learning flood dynamics prediction model~\cite{qian2019physics} which is a substitute of the PDE dynamics model for real-time implementation purpose.

\subsection{Real-time mobile sensor task scheduling algorithm for large scale systems}

% \begin{figure*}
%         \centering
%         \begin{subfigure}[b]{0.225\textwidth}
%             \centering
%             \includegraphics[width=\textwidth]{figs/ex1struc.pdf}
%             \caption[]%
%             {{\small }}    
%             \label{fig:mean and std of net14}
%         \end{subfigure}
%         \hfill
%         \begin{subfigure}[b]{0.75\textwidth}  
%             \centering 
%             \includegraphics[width=\textwidth]{figs/ex1sequence.pdf}
%             \caption[]%
%             {{\small }}    
%             \label{fig:mean and std of net24}
%         \end{subfigure}
%         \vskip\baselineskip

%         \caption[ The illustration of an example with action constraints. ]
%         {\small The illustration of an example with action constraints. } 
%         \label{fig:example1}
% \end{figure*}

\begin{figure*}
        \centering
        \begin{subfigure}[b]{0.325\textwidth}
            \centering
            \includegraphics[width=\textwidth]{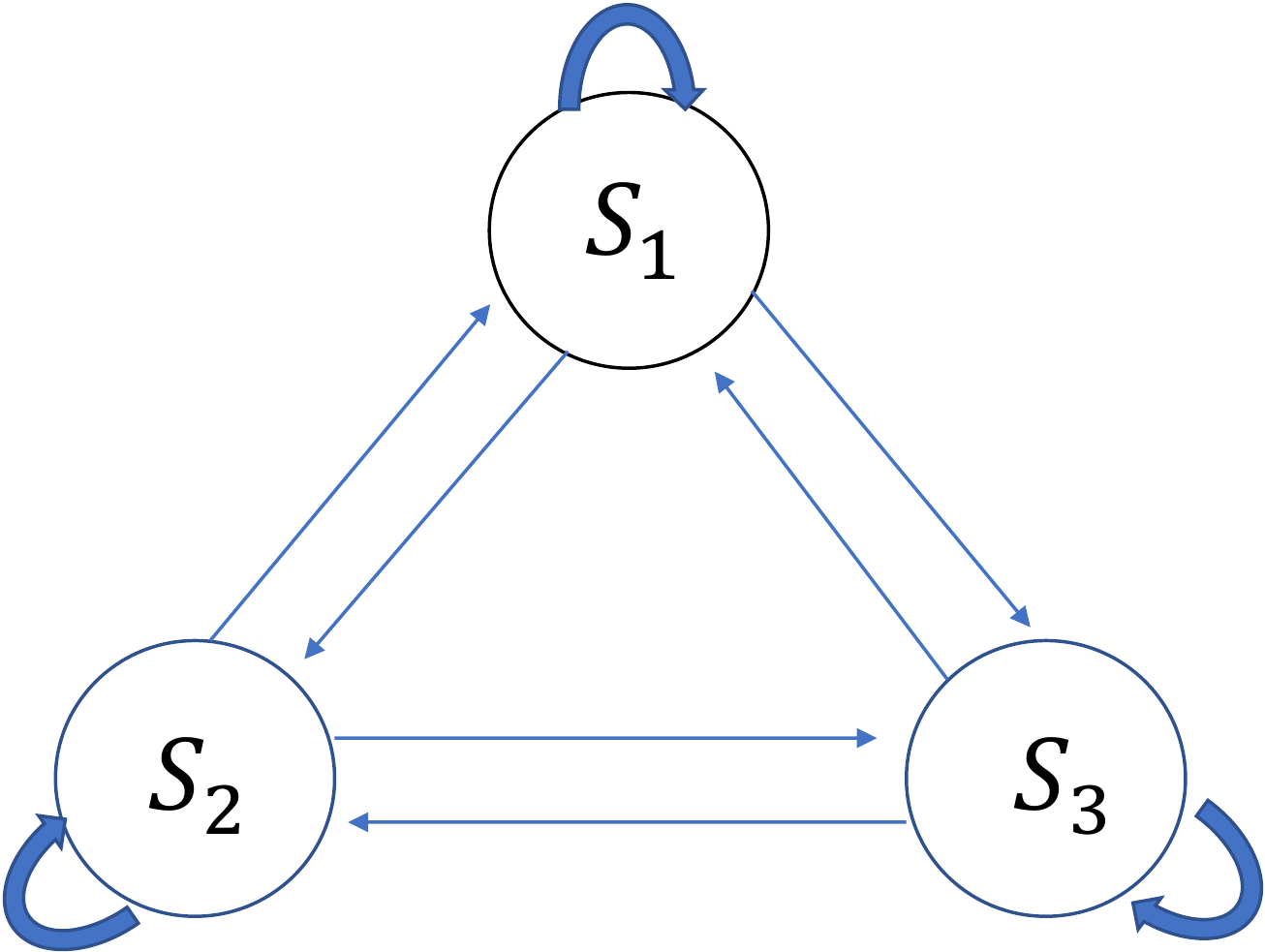}
            \caption[]%
            {{\small }}    
            \label{fig:mean and std of net14}
        \end{subfigure}
        \hfill
        \begin{subfigure}[b]{0.6\textwidth}  
            \centering 
            \includegraphics[width=\textwidth]{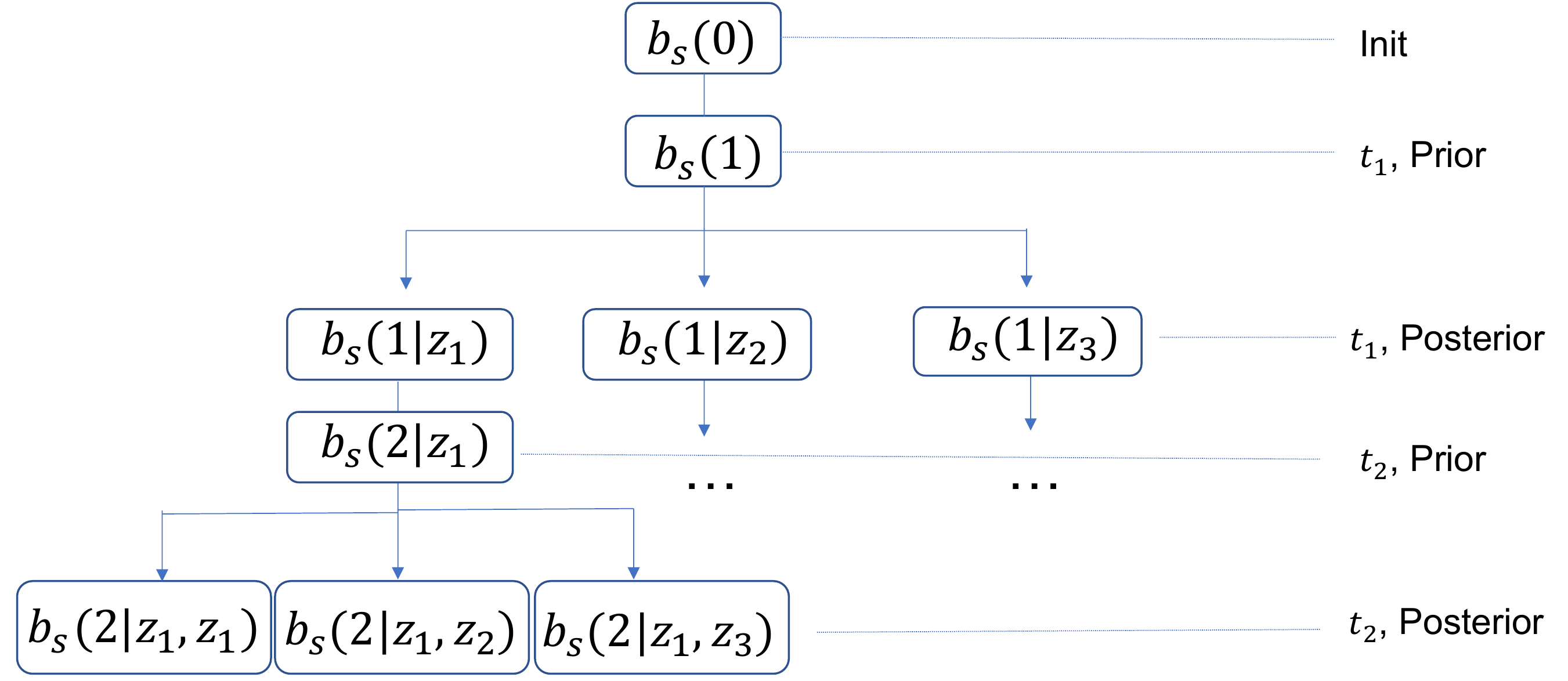}
            \caption[]%
            {{\small }}    
            \label{fig:mean and std of net24}
        \end{subfigure}
        \vskip\baselineskip

        \caption[]
        {\small Illustration of a small-scale monitoring setup. (a) shows a simple system with three states $S_{1, 2, 3}$. It also shows the paths from which mobile sensors can travel between states to observe at each step. (b) constructs the full search tree which describes all possible action sequences of a single mobile sensor monitoring this system.} 
        \label{fig:example2}
\end{figure*}

Without loss of generality, let us consider a simple example to illustrate how the algorithm simplifies a full forward search tree. Assuming that one mobile agent carrying sensors can move among locations and obtain a measurement at each step in subplot (a) of Figure~\ref{fig:example2}, a forward search tree is constructed in the (b) subplot of Figure~\ref{fig:example2} based on the possible sequences of sensing locations. Two factors contribute to a large number of computations. First, the number of branches in the tree to be searched grows exponentially. To be concrete, to predict $T$ steps ahead, the number of branches to be searched is of the order $\mathcal{O}(D_{obs}^T)$ in which $D_{obs}$ is the dimensionality of the candidate set for observations. Second, each data assimilation step is time-consuming. For instance, the approximated computational complexity of the Ensemble Kalman filter (EnKF) is $\mathcal{O}(N D^2)$ where $D$ is the dimension of the system and $N$ is the number of ensembles. Combining the above two factors, the overall computational complexity of the forward search tree is $\mathcal{O}(D_{obs}^{T+1}( T N D^2 + \mathcal{O}_{p}))$ in which $\mathcal{O}_{p}$ is the computational complexity of model prediction. The computation cannot be implemented in real-time because $D$ would be a large number for large scale environmental systems. Branch pruning methods~\cite{atanasov2014information} can not reduce the computational amount significantly while preserving the non-myopic planning feature when both $D$ and $D_{obs}$ are large. 

% In this paper, the search tree is approximated such that most of the computationally heavy data assimilation step is get rid of during the forward planning. A well designed function is added to make up the missed information caused by skipping data assimilation steps.

\begin{figure}
    \centering
    \includegraphics[width = 0.7\textwidth]{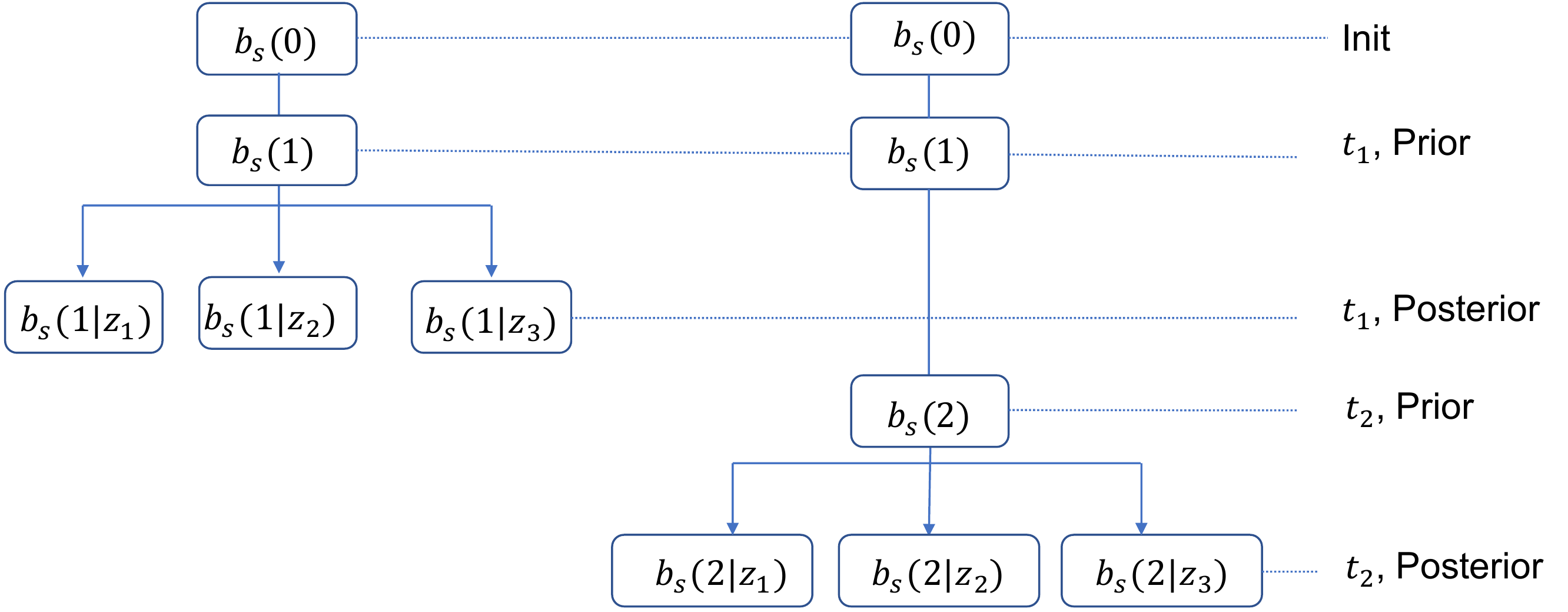}
    \caption{Approximated Search Tree. This approximation omits the data assimilation steps shown in Figure~\ref{fig:example2}(b). The data assimilation steps are only conducted in the last step of each separate tree to compute the rewards value. The function $\Gamma$ connects rewards obtained at different time steps together to form a new search tree with less computational complexity.}
    \label{fig:exsimplify}
\end{figure}

To significantly reduce the computational complexity, the original search tree shown in Figure~\ref{fig:example2} is decomposed into $T$ independent prediction trees. The approximation shown in Figure~\ref{fig:exsimplify} omits the data-assimilation steps except at the last step of each tree. The model prediction power is fully preserved and the data assimilation step is kept in the last step of each decomposed tree for the purpose of calculating rewards. However, this simplification loses the causal relations among a sequence of observation actions. To address this, an empirical function is applied. The correlation between observation sets $\omega_i$ and $\omega_j$ is denoted as $\Sigma(\omega_i, \omega_j)$. We multiply a coefficient $\Gamma(\omega_{t})$ computed from Equation~\ref{equ:coef} with the reward gained from observing $\omega_{t}$ at time step $t$. This coefficient serves to model the causal relationship between the previous observations and new observation set $\omega_t$. The computational complexity of the simplified algorithm is $\mathcal{O}(D_{obs} T N D^2 + NT\mathcal{O}_{p} +  D_{obs}^{T+1}\mathcal{O}_{\Gamma})$ in which $\mathcal{O}_{\Gamma}$ is the computational complexity of computing the function $\Gamma(\omega)$ for a single time. The term $D_{obs}^{T+1}\mathcal{O}_{\Gamma}$ may dominate under the large $T$ situation. However, in this paper, because $\mathcal{O}_{\Gamma}$ is much smaller than $N D^2$ and we will not set $T$ to be too huge, $D_{obs}^{T+1}\mathcal{O}_{\Gamma}$ is not a computational heavy burden. Also, the computational complexity of model prediction $\mathcal{O}_{p}$ will not be an obstacle to real-time implementation because of the deep learning prediction model. The comparison of the computational complexity between the full branch search tree and the simplified one is shown in Table~\ref{tab:time_compare}. We fix some specific values for forward prediction horizon $T$, system dimension $D$ and observation locations $D_{obs}$ to compare the computational time for full search tree and simplified search method. It can be observed that when the dimension of $D$ is $1000$ and $D_{obs}$ is $50$, the full search tree based method cannot handle the computational amount in real time any more.

\begin{table}[]
\caption{Computational time comparison between the simplified method and the full search tree}
\begin{center}
\begin{tabular}{c|c|c|c|c}
\hline
\begin{tabular}[c]{@{}c@{}}Prediction\\ Horizon\end{tabular} & \begin{tabular}[c]{@{}c@{}}Dimension\\ of D\end{tabular} & \begin{tabular}[c]{@{}c@{}}Dimension\\ of $D_{obs}$\end{tabular} & \begin{tabular}[c]{@{}c@{}}Full Search Tree\\ Time\end{tabular} & Simplified Time \\ \hline
\multirow{3}{*}{$T = 2$}                                     & 100                                                      & 10                                                               & 0.06s                                                           & 0.1s            \\ \cline{2-5} 
                                                             & 1000                                                     & 50                                                               & 51s                                                             & 7s              \\ \cline{2-5} 
                                                             & 10000                                                    & 100                                                              & $>1h$                                                           & 420s            \\ \hline
\multirow{3}{*}{$T = 3$}                                     & 100                                                      & 10                                                               & 0.7s                                                            & 1.5s            \\ \cline{2-5} 
                                                             & 1000                                                     & 50                                                               & 2600s                                                           & 28s             \\ \cline{2-5} 
                                                             & 10000                                                    & 100                                                              & $>1$ day                                                        & 2600s           \\ \hline
\multicolumn{3}{c|}{Computational Complexity}                                                                                                                                              & $ \mathcal{O}(D_{obs}^{T+1}( T N D^2)) $                                                             & $ \mathcal{O}(D_{obs} T N D^2 +  D_{obs}^{T+1}\mathcal{O}_{\Gamma}) $             \\ \hline
\end{tabular}
\end{center}
\label{tab:time_compare}
\end{table}

The expression of $\Gamma(\omega)$ is shown in Equation~\ref{equ:coef}. We need $f_e(x)$ to be high when $x$ is small. Indeed, the more we gather information on specific states, the less uncertain we are about these states and other states that are strongly correlated. The correlation between states is based on empirical correlation drawn from a large amount of simulations. Detailed discussion about $f_e(x)$ is in Section~\ref{sec:experiments}.

\begin{equation}\label{equ:coef}
    \Gamma(\omega_{t+1}) = f_{e}((\Pi_{i = 1}^{t} \Sigma(\omega_i, \omega_{t+1})\exp{(i-t-1)}) + 0.01
\end{equation}

% Recent work~\cite{chen2019autonomous} has involved neural networks to directly map from system and sensor states to future actions. However, this paper argues that producing enough training data is so computationally demanding that it is not realistic for large scale systems.

% From equation~\ref{equ:klgaussian}, we can see the KL divergence between two distributions depend on their difference in mean value and second momentum information. For the mean value part, we depend on the expected predicted value to measure the future expected quality of a sensing location. With this assumption, we can notice that the expected values are generally unchanged before and after the data-assimilation. Thus, it doesn't contribute to the rewards. For the terms in rewards contributed by second order momentum of the distribution, the main issue caused by such an approximation is that we may repeatedly guide our sensors to a specific states or a group of states which are strongly correlated. Because in the original computation tree, once a state is measured, the uncertainty thus rewards related to those strongly related states will be reduced in future time steps. Approximation, we will miss this part of information when planning. 

The approximation is summarized in Equations~\ref{equ:MSEerror_approx}~\ref{equ:dynamicsconstraint_approx}~\ref{equ:obs_approx}~\ref{equ:assimilation_approx}~\ref{equ:reachableconstraint_approx}, and the corresponding algorithm is presented in algorithm~\ref{alg:sparse}.

\begin{equation}
\label{equ:MSEerror_approx}
\min_{\mathbf{\omega}(t_0), \dots, \mathbf{\omega}(t_0 + T)}\frac{1}{T} \sum_{t=t_{0}}^{T + t_{0}} \gamma^{t - t_0} \Gamma(\omega(t)) \mathbf{R}(\omega(t), b_{S}(t|Z_{1:t_0-1}, \hat{Z}_{\omega(t)}))
\end{equation}
Subject to
\begin{equation}
\label{equ:dynamicsconstraint_approx}
b_{S}(t|Z_{1:t_0}) = f(b_{S}(t-1|Z_{1:t_0}), u(t))
\end{equation}
\begin{equation}
\label{equ:obs_approx}
    \hat{Z}_{\omega(t)} = H_{\omega(t)}(b_{S}(t|Z_{1:t_0-1}))
\end{equation}
\begin{equation}
\label{equ:assimilation_approx}
b_{S}(t|Z_{1:t_0}, \hat{Z}_{\omega(t)}) = M(b_{S}(t|Z_{1:t_0}), \hat{Z}_{\omega(t)})
\end{equation}
\begin{equation}
\label{equ:reachableconstraint_approx}
\text{ s.t. } \omega(t+1) \in \mathcal{A}(\omega(t))
\end{equation}

% ================================= Begin of Alg. 2 ========================== % 
\begin{algorithm}
  \caption{Fast Sensor task Scheduling Algorithm for single mobile sensor}\label{alg:sparse}
  \begin{algorithmic}[1]
  \State \textbf{Input:} $\Omega(t_0)$, $\dots$, $\Omega(t_0 + k)$; $b_S(t|Z_{1:t_0})$, $k$, $u(t_0 + 1)$, $\dots$, $u(t_0 + k)$, $T$; \Comment{$\Omega(t)$ is the set of already fixed observation locations at time step $t$.}
  \State \textbf{Init:} $t = 1$;

      \While{$t\leq T$}\Comment{Prediction $T$ steps}
        \State $b_S(t + 1|Z_{1:t_0}) = f(b_S(t + 1|Z_{1:t_0}), u(t+1))$; 
        \State t += 1 
      \EndWhile\label{predictionwhile}
      \State \textbf{set }$t = 1$
      \While{$t\leq T$}\Comment{Compute Rewards for each future step}
      \For{$\omega_i(t)\in \mathcal{A}(t)$}\Comment{$\mathcal{A}(t)$ is the set of all possible observation locations at time $t$}
      
        \State \textbf{Compute: }$\mathbf{R}(\omega_i(t), b_S(t |Z_{1:t_0}, \hat{Z}_{\omega(t)}))$
        \EndFor
        \State \textbf{Normalize: }$\forall \omega_i(t) \in \mathcal{A}(t), \mathbf{R}(\omega_i(t), b_S(t |Z_{1:t_0}, \hat{Z}_{\omega(t)})) = \frac{\mathbf{R}(\omega_i(t), b_S(t |Z_{1:t_0}, \hat{Z}_{\omega(t)}))}{\max_{\omega_j(t)\in \mathcal{A}(t)} (\mathbf{R}(\omega_i(t), b_S(t |Z_{1:t_0}, \hat{Z}_{\omega(t)})))}$
        \State $t += 1$
      \EndWhile

      \State Organize all the possible sequences of observations $Seq_i$ into set $Seq$
      
      \State Compute the value $\sum_{t=t_{0}}^{T} \gamma^{t - t_0} \Gamma(\omega(t)) \mathbf{R}(\omega(t), b_{S}(t|Z_{1:t_0-1}, \hat{Z}_{\omega(t)}))$ for each $seq_i$
      
      \State \textbf{return:} $\max Seq_{i}$
  \end{algorithmic}
\end{algorithm}
% ================================= End of Alg. 2 ========================== % 

% As we target at monitoring a large scale of distributed parameters system, we can generally divide the large system into several small sub-systems such that the states between each subsystems can be assumed not correlated and the states within a subsystem can be assumed to be correlated. This may serve as an explanation that why we can set the rule that in the prediction process, we construct the graph such that we don't consucutively visit the same local area.

% (add a figure here)

\subsection{Ensemble Kalman filter(EnKF)}
{Data assimilation is one of the main components in the frame shown in Figure~\ref{fig:generalflow}. As mentioned in Section~\ref{sec:ltrreview}, the proposed framework can be deployed along with general data assimilation methods. Meanwhile, since this framework targets at city-scale distributed parameters systems, we apply the Ensemble Kalman Filter (EnKF) for assimilation.} EnKF has been shown to be a computationally efficient and robust data assimilation method~\cite{evensen1994sequential, houtekamer1998data, hamill2001distance} for large scale geographical and environmental applications. With the assumption of Gaussian Distribution, EnKF follows the same format as the classical Kalman filter to conduct data assimilation and employs Monte Carlo methods to capture nonlinearities. The ensemble of estimated state $b_S$ is denoted as $B_S$. The measurement update of EnKF is introduced in Equation~\ref{equ:enkf}. $\Sigma(B_S(t|Z_{1:t_0}))$ is the covariance matrix of $B_S(t|Z_{1:t_0})$ and $R$ is covariance matrix of measurement noise induced by $\omega(t)$. 
\begin{equation}
\label{equ:enkf}
    B_S(t|Z_{1:t_0}, Z_t) = B_S(t|Z_{1:t_0}) + \Sigma(B_S(t|Z_{1:t_0})) H_{\omega(t)}^T K_{\omega(t)} (Z_t - H_{\omega(t)} B_S(t|Z_{1:t_0}))
\end{equation}
in which
\begin{equation*}
    K_{\omega(t)} = (H_{\omega(t)} \Sigma(B_S(t|Z_{1:t_0})) H_{\omega(t)}^T + R)^{-1}
\end{equation*}

% Several techniques are applied to improve the performance of EnKF in this work. For example, a pair of independently generated ensembles $\{ B^1_S, B^2_S \}$ are used alternatively to track the belief state~\cite{houtekamer1998data}.

\subsection{Data-Driven Dynamics Model}
This work depends on the deep learning based data-driven model invented in~\cite{qian2019physics} to predict flood dynamics in real-time. The data-driven model is derived from training a Convolutional Neural Network(CNN) fed with data simulated by a Shallow Water Equation(SWE) solver.  Table~\ref{tab:speed} compares the time taken to compute dynamics by data-driven model and PDE solver. The data-driven model not only boosts the prediction but also keeps the precision. It was validated that the prediction by the data-driven model will not diverge from predictions by the SWE solver when evolving along with time. In Section~\ref{sec:experiments} of this paper, it is validated with experiments that Ensemble Kalman Filter could have good performance in assimilating measurement data with the data-driven prediction model.

\begin{table}
\centering
\begin{tabular}{|l|l|} 
\hline
Methods   &       Computation Speed            \\ 
\hline
PDE Solver  &       $1 \times$           \\ 
\hline
Model 1   &        $50000 \times$           \\ 
\hline
\end{tabular}
\caption{Computational time improvements. All computations are performed on a Intel i7-7700 CPU $@$3.60GHz. Since the computational time of the PDE solver depends on the initial conditions and the inputs of the problem, we average the PDE solver performance over a large number of different cases.~\cite{qian2019physics}}
\label{tab:speed}
\end{table}

\subsection{Reward Function}

Information theory~\cite{shannon1948mathematical} provides ways to measure the uncertainty of a system. Entropy describes the uncertainty of the belief of a system. The reduction of entropy through measurement can be used as a way to measure information gain~\cite{hero2011sensor}. The trace or log of determinant of the covariance matrix is also employed by many works~\cite{vitus2012efficient, jawaid2015submodularity, schlotfeldt2018anytime, atanasov2014information} to measure the uncertainty of systems. From the estimation point of view, Cramer Rao bound~\cite{tichavsky1998posterior} is the lower bound of error for nonlinear estimators. It is also reasonable to employ it as the rewards function for sensor actions~\cite{shen2014sensor, hernandez2004multisensor}. In many other works about sensor management~\cite{hero2007foundations, ristic2010sensor, chong2009partially}, Kullback Divergence(KL Divergence)~\cite{kullback1951information} or Renyi Divergence are popular choices. 

\subsection{Real-time solution to multiple sensors case}
The dimensionality of observation space is exponential in the number of sensors $n$ if we follow algorithm~\ref{alg:sparse}. Real-time implementations are only possible when $n$ is small. \cite{krause2008near} introduced a greedy algorithm to select locations for multiple fixed sensors. This greedy algorithm is built upon the submodular property of the mutual information reward function. \cite{schlotfeldt2018anytime} relies on coordinate descent which is a similar greedy algorithm with~\cite{krause2008near} to schedule tasks for a number of mobile sensors. To apply the coordinate descent algorithm to multiple mobile sensors, we need to fix an order of sensors to be computed in advance. Algorithm~\ref{alg:alg-multiple} extends algorithm~\ref{alg:sparse} and follows a similar coordinate descent style as introduced in~\cite{schlotfeldt2018anytime}.

\begin{algorithm}
  \caption{Multiple Mobile Sensor task Scheduling}\label{alg:alg-multiple}
  \begin{algorithmic}[3]
  \State \textbf{Input:} $b_S(t|Z_{1:t_0})$, $k$, $u(t_0 + 1)$, $\dots$, $u(t_0 + k)$, $n_S$; \Comment{$n_S$ is the total number of mobile sensors}
  \State \textbf{Init:} $\Omega(t_0) = \{\}$, $\dots$, $\Omega(t_0+k) =  \{ \}$ \Comment{$\Omega(i)$ is the set to save all the locations decided to be visited at time step $i$}
  \For{$i = 1:n_S:$}
    \State $(\omega(t_0), \dots, \omega(t_0 + k)) =$ \text{Fast Sensor task Scheduling Algorithm for single sensor}($\Omega(t_0)$, $\dots$, $\Omega(t_0 + k)$, $b_S(t|Z_{1:t_0})$, $k$, $u(t_0 + 1)$, $\dots$, $u(t_0 + k)$)
    \State $\Omega(t_0) = \{ \Omega(t_0), \omega(t_0) \}$, $\dots$, $\Omega(t_0 + k) = \{ \Omega(t_0 + k), \omega(t_0 + k) \}$
  \EndFor
  \State \textbf{Return: }$\Omega(t_0)$ \Comment{$\Omega(t_0)$ is the set of locations to be observed at next step}
  \end{algorithmic}
\end{algorithm}
\section{Numerical Experiments}
\label{sec:experiments}

\begin{figure}
    \centering
    \includegraphics[width =0.9\textwidth]{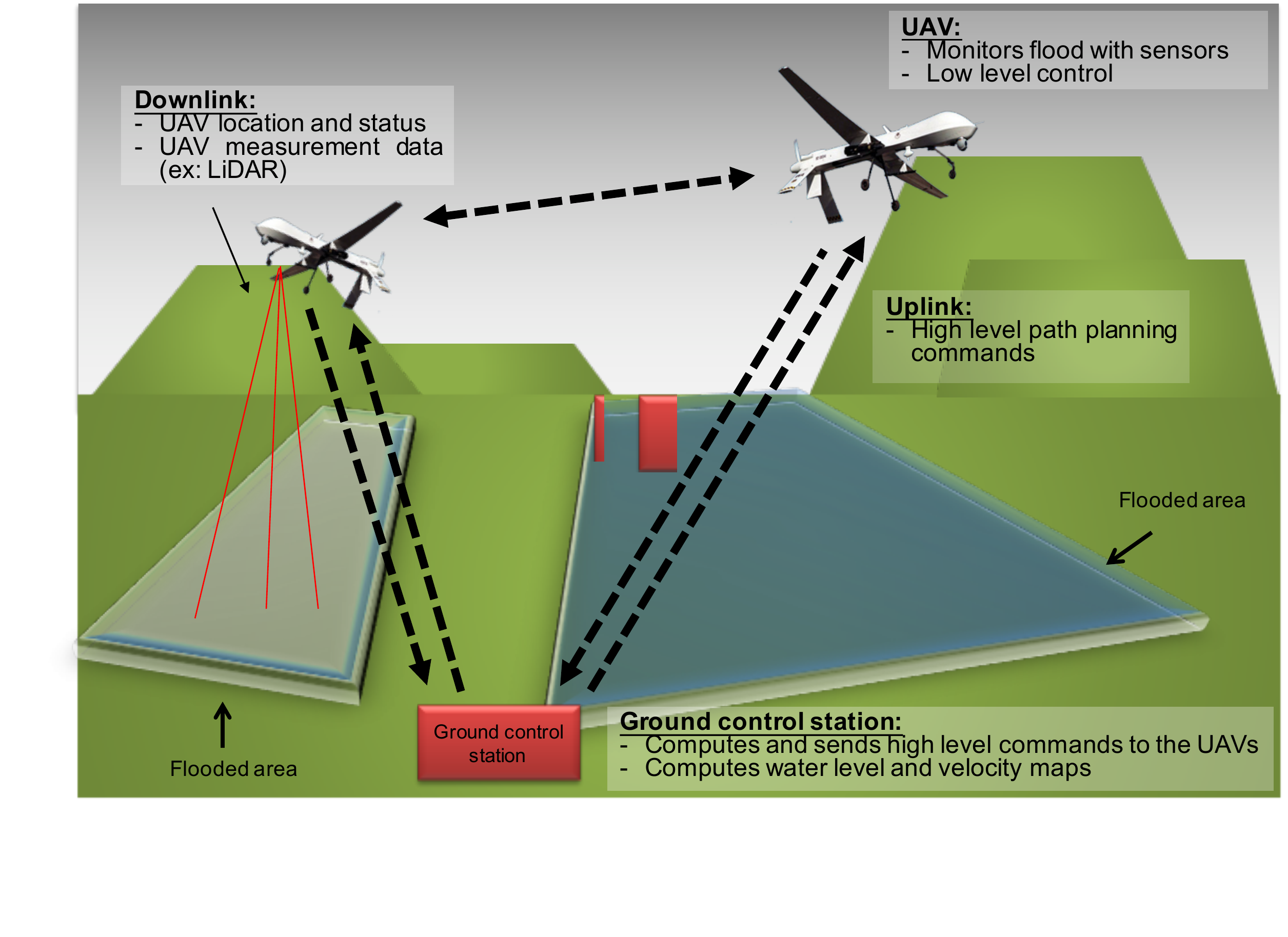}
    \caption{Illustration of the UAV flood monitoring procedure. UAVs carrying sensors are distributed around the area to be monitored. Ground control stations send commands to UAVs with high level task commands (target locations). This paper focuses on the algorithm to generate high level tasks commands, while an earlier paper focused on low-level UAV control~\cite{eldad2016real}.}
    \label{fig:illustration}
\end{figure}

Section~\ref{sec:experiments} analyzes the performance of the real-time mobile sensors task scheduling algorithm with experiments. Since floods are events that occur over large (city-scale) spatial domains, a very large number of sensors operating for a few hours is required to accurately monitor these events, which have not been deployed in a given location to date. There is thus a lack of accurate, a city-scale flood propagation in earlier literature. To adddress this, we use simulated flood data based on~\cite{qian2019physics} to conduct numerical experiments. Austin, Texas is chosen as the studied area shown in Figure~\ref{fig:illustration}. Using Austin as an example does not restrict the application domain since the method can be applied to any other city where the flood would follow the Shallow Water Equation~\cite{qian2019physics}, with different model parameters. As a city with numerous unmonitored creeks, Austin would benefit from such a mobile flood tracking system. Furthermore, ~\cite{qian2019physics} provides a data-driven model to boost the computational speed of the forecast step, which is key to real-time implementation of the present path-planning framework.

Floods caused by heavy rains are simulated for mobile sensors to track. It is assumed that the Partial Differential Equation solver simulates trues states of the environment. The planning frame relies on the deep learning model~\cite{qian2019physics} to predict the development of floods and has no access to any of the true rain inputs or flood states but only noisy values corrupted by Gaussian noise from predictions and sensor measurements. In this work, the rewards are defined using the log-determinant of the covariance matrix. The reward at each step is defined by Equation~\ref{equ:mutualinfo}.
\begin{equation}
    \label{equ:mutualinfo}
    \mathbf{R}(\omega(t), b_{S}) =
    \log \det(\Sigma(b_{S}(t)|\hat{Z}_{\omega(t)}))
\end{equation}

\begin{figure*}
    \centering
    \includegraphics[width =0.6 \textwidth]{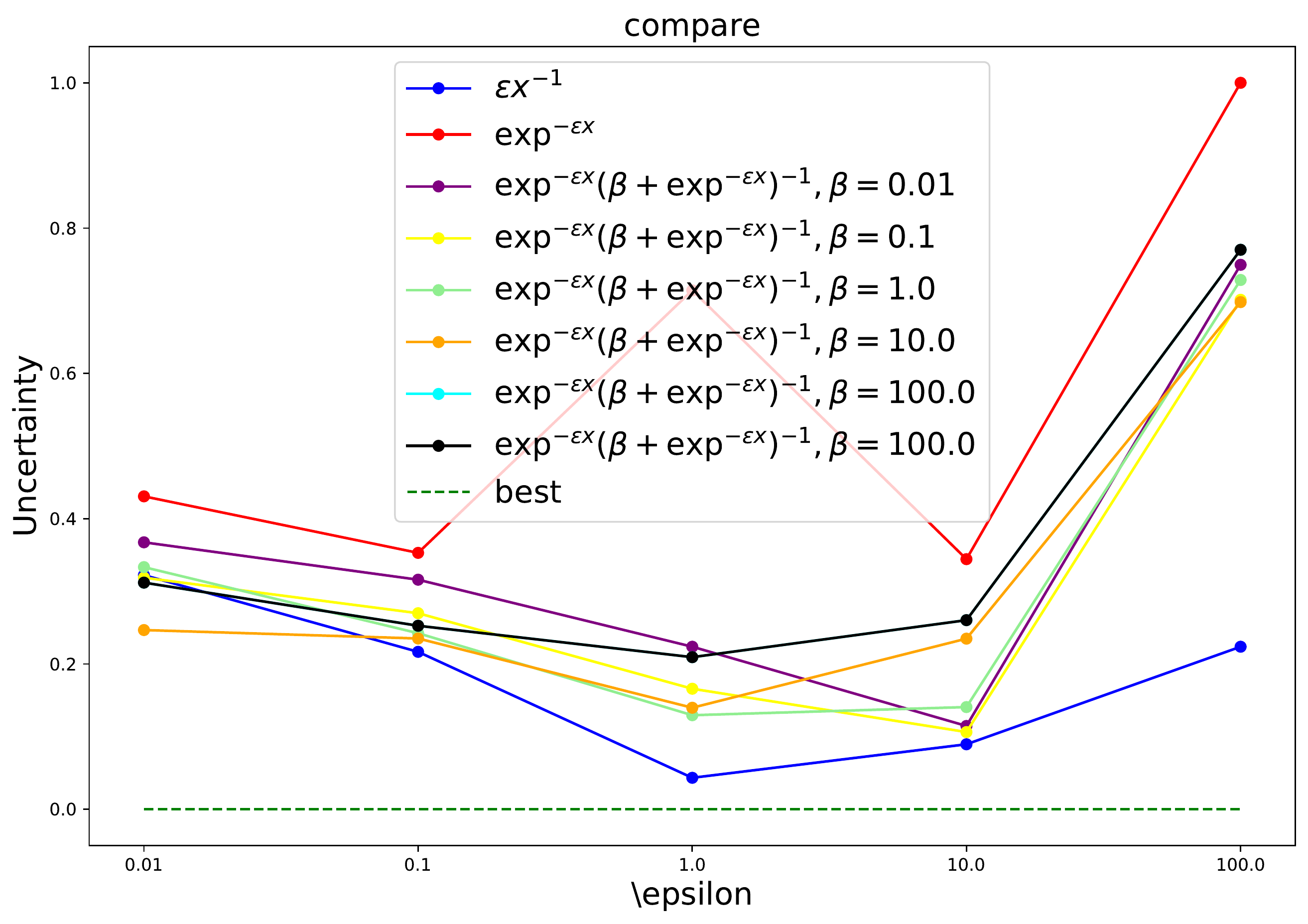}
    \caption{Comparison of different hyper-parameters and empirical function in $\Gamma(\omega)$. The experiments are conducted with 30 scenarios of each a different set of observation candidates is considered. In order to reduce the computation time, the chose a small set of observation candidates.}
    \label{fig:ablationstudy}
\end{figure*}

\begin{figure*}
    \centering
    \includegraphics[width = \textwidth]{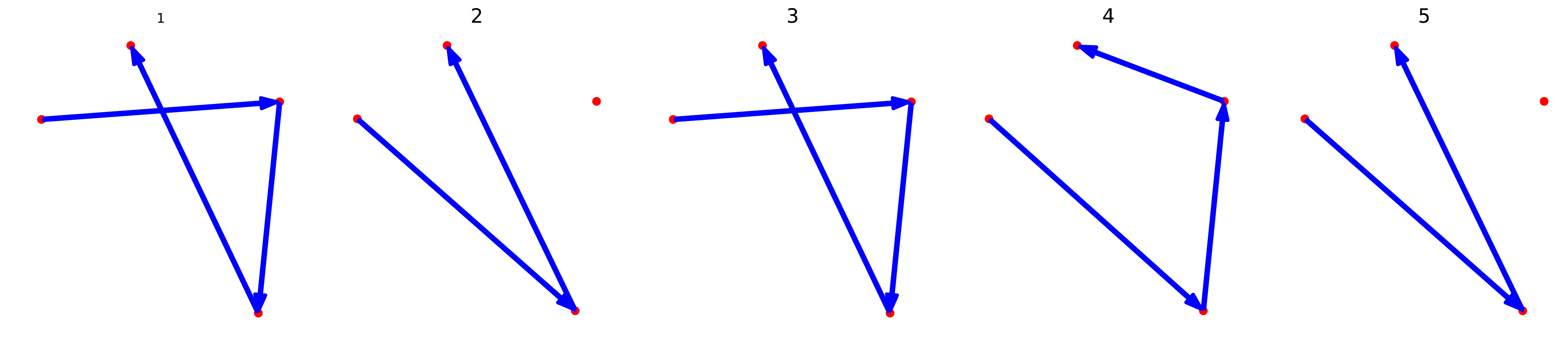}
    \caption{Comparison between result by the full forward search tree and results by approximated search tree. 1 shows the optimal sensor task sequence computed by full forward search tree. 2 shows the observation sequence searched by approximated search tree without adding function $\Gamma(\omega(t))$. 3 shows the result by setting $f_e(x) = 1.0/x$. 4 shows the result by setting $f_e(x) = \exp^{-10.0 x}$. 5 shows the result by setting $f_e(x) = \frac{\exp^{-100.0 x}}{1.0 + \exp^{-100.0 x}}$.}
    \label{fig:track_compare}
\end{figure*}

To evaluate the performance of $\Gamma(\omega_t)$ and search for the best hyperparameters, simple experiments are conducted to compare the result by full tree search and results by approximated search tree with different heuristic functions. In order to make the computation tractable for the full search case, each experiment selects only $4$ different locations as candidate measurement locations for one mobile sensor. The planning horizon is set as $4$. In Figure~\ref{fig:ablationstudy}, the ablation study is conducted with $30$ different observation candidates sets. The ablation study indicates that with an appropriate function $\Gamma(\omega(t))$, the algorithm can find observation sequences performing almost as well as sequences computed by full branch search tree while requiring considerably less computational time according to Table~\ref{tab:time_compare}. To qualitatively understand the performance, according to the Figure~\ref{fig:track_compare}, we can see that without $\Gamma(\omega(t))$, the sensor will be guided to visit the same spot again and again continuously. With an appropriate choice of $\Gamma(\omega(t))$, the mobile sensor can be guided to visit different spots at different steps. 

The performance of algorithm~\ref{alg:sparse} is studied with two experiments. The first experiment compares the performance of a single mobile sensor under myopic and non-myopic strategies. The second experiment assumes few mobile sensors working together with a number of fixed sensors installed along the river. We analyze quantitatively the performance of the algorithm~\ref{alg:sparse} with different prediction horizon lengths and show qualitatively the estimation results. Experiments with a large number of mobile sensors are conducted to show that following algorithm~\ref{alg:alg-multiple} could achieve much better performances in managing a large number of mobile sensors than other strategies. To evaluate the performance, we use $30$ different pieces of time series data and generate $10$ different initial guess of states for each time-series data. Notice that even with algorithm~\ref{alg:sparse} and algorithm~\ref{alg:alg-multiple}, the computation amount is still huge for large scale systems.
% The first two experiments limit the amount of mobile sensors as we would like to explicitly compare the performance of the algorithm under different settings.

\subsection{Single Mobile Sensor}

In the first experiment, we follow the algorithm~\ref{alg:sparse} to schedule the sensing task for a single mobile sensor. Although single mobile sensor is a particular case of the multi-agent problem, this experiment shows with the simple setting that the proposed algorithm can guide mobile sensors to choose more informative ways to explore and overcome the drawbacks of myopic strategies. We compare the performance of myopic and non-myopic planning strategies. With the non-myopic strategy, the prediction horizon is set to be 3 steps ahead. Monte Carlo simulations are used to show that the informative scheduling algorithm achieves near-optimal performance.

\begin{figure*}
        \centering
        \begin{subfigure}[b]{0.4\textwidth}
            \centering
            \includegraphics[width=\textwidth]{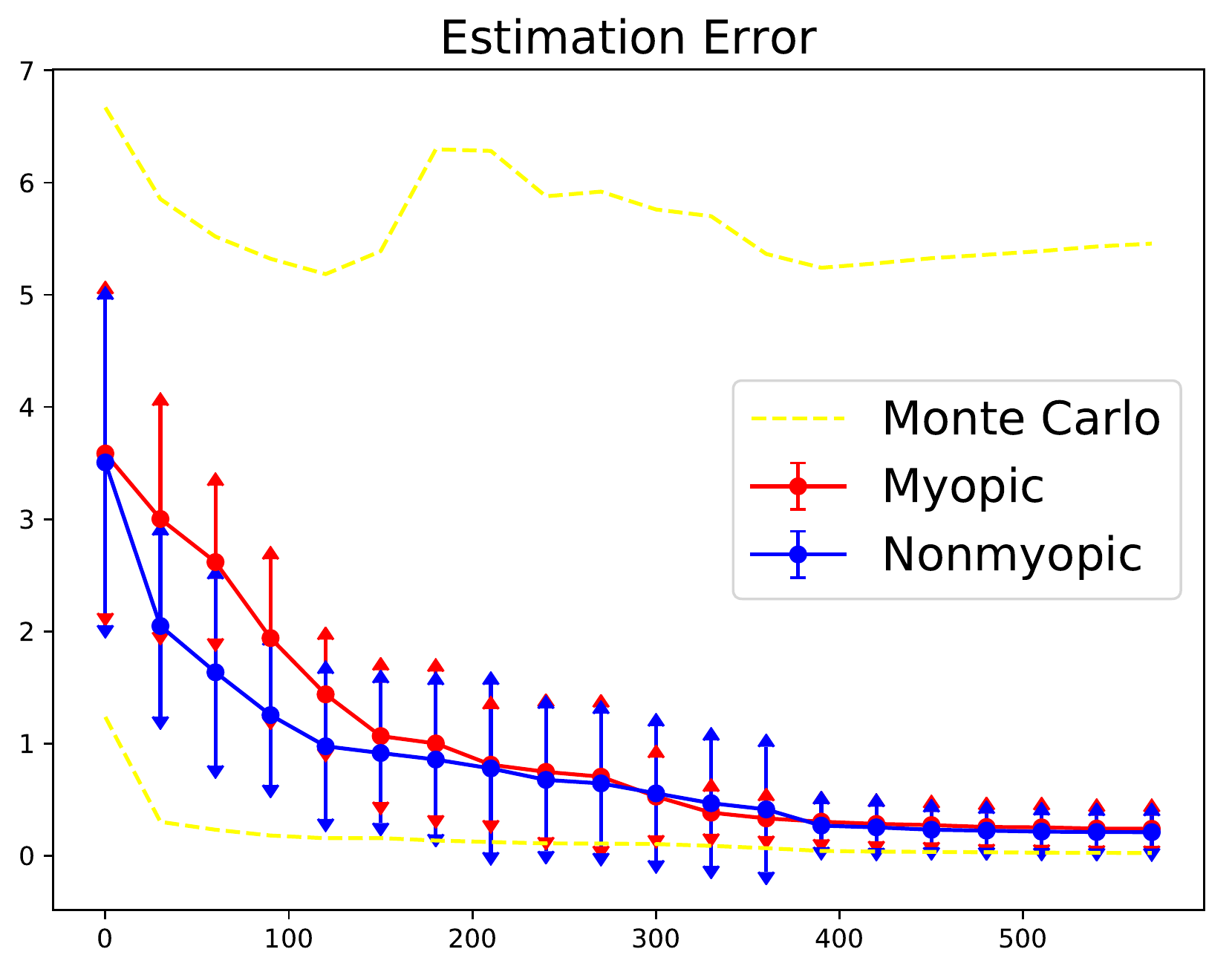}
            \caption[]%
            {{\small }}    
            \label{fig:mean and std of net14}
        \end{subfigure}
        \quad
        \begin{subfigure}[b]{0.4\textwidth}  
            \centering 
            \includegraphics[width=\textwidth]{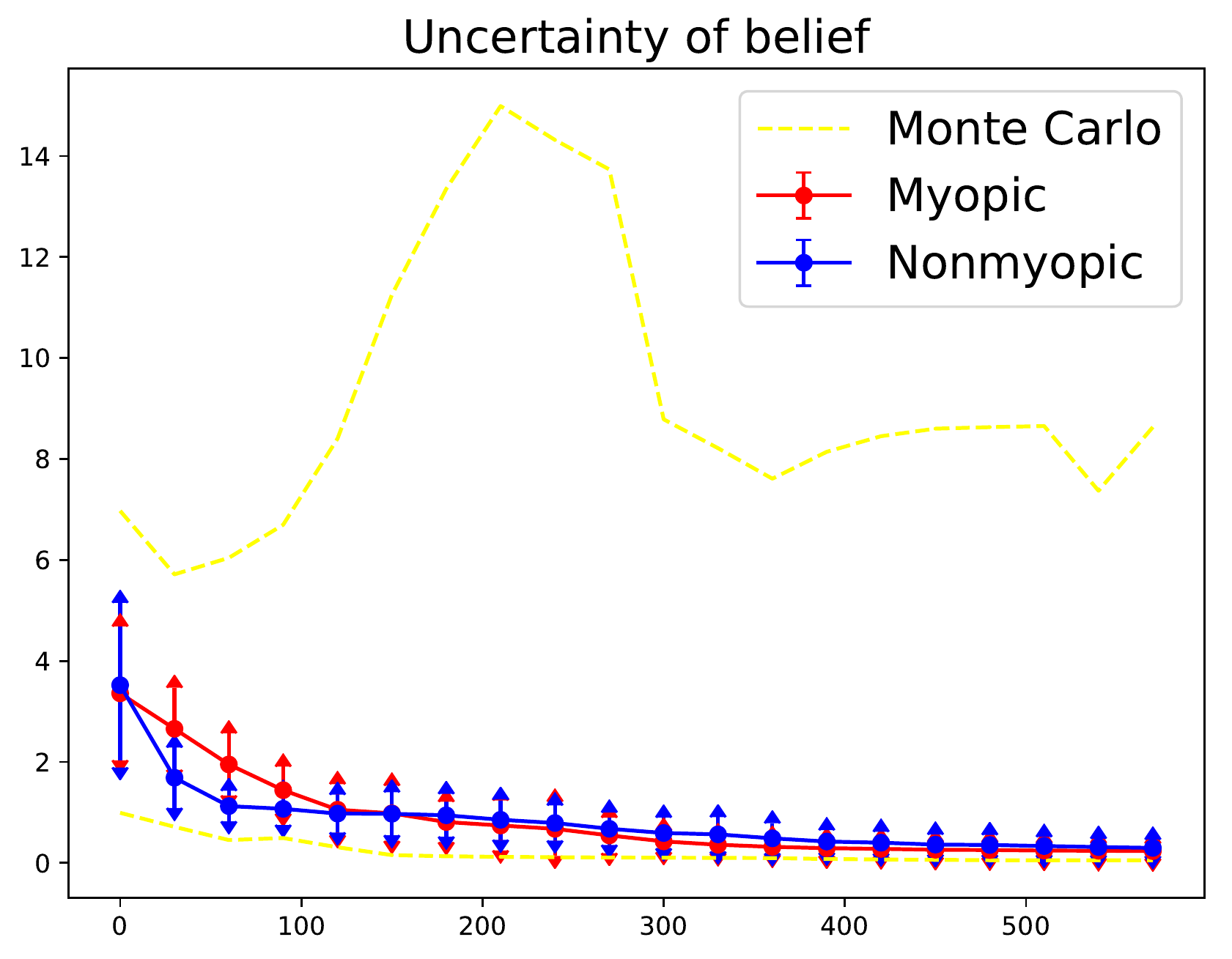}
            \caption[]%
            {{\small }}    
            \label{fig:mean and std of net24}
        \end{subfigure}
        % \vskip\baselineskip
        % \begin{subfigure}[b]{0.3\textwidth}   
        %     \centering 
        %     \includegraphics[width=\textwidth]{figs/exp2_infogain.pdf}
        %     \caption[]%
        %     {{\small }}    
        %     \label{fig:mean and std of net34}
        % \end{subfigure}
        \quad

        \caption[Performance comparison between Nonmyopic and Myopic strategies with a single mobile sensor. ]
        {\small Performance comparison between Nonmyopic and Myopic strategies with single mobile sensor. The results are averaged over 300 cases that come from 30 different simulated time series data and 10 different initial guesses for each. Results are combined together from each experiment after normalization. The vertical lines in the plot indicate the variance of performance through different cases. Monte-Carlo Simulation produces the range of performance for all 300 cases. (a) shows the estimation error. (b) show the uncertainty of belief states.} 
        \label{fig:single}
\end{figure*}

According to Figure~\ref{fig:single}, we can see that mobile sensors planned using algorithm~\ref{alg:sparse} would reduce the estimation error and the uncertainty of the system in time. Besides, the non-myopic planning strategy helps improve the performance over myopic strategy. This is due to the fact that non-myopic strategies tend to bring the mobile sensor to explore new areas such that information from different areas can be collected. Comparing to the cases simulated by Monte-Carlo methods, both strategies' performance is close to optimal performance after a period.

\begin{figure*}
    \centering
    \includegraphics[width = \textwidth]{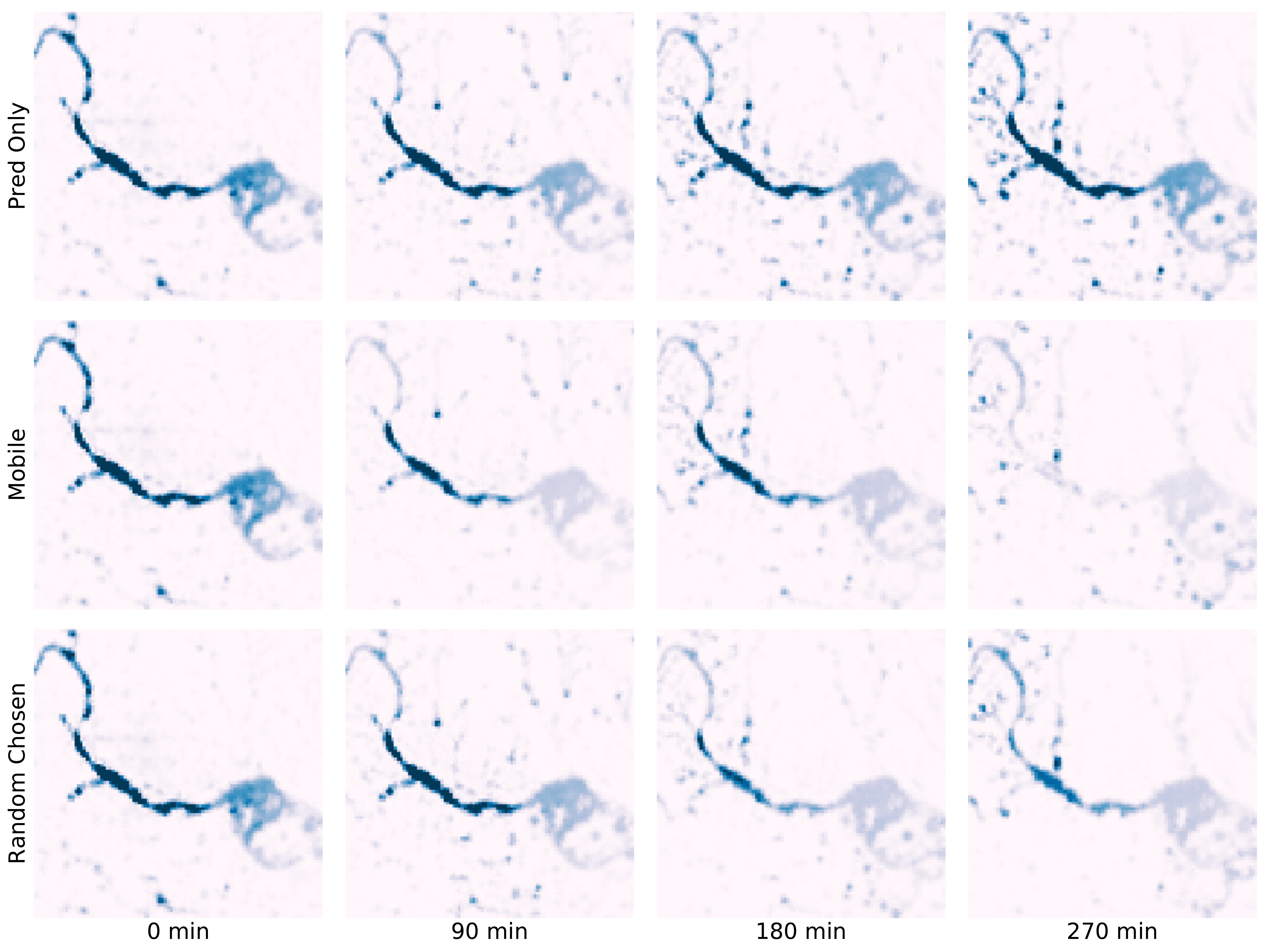}
    \caption{Estimation error by different sensor strategy. The difference in depth of water with real status is represented by a color shade. The first row is obtained by only predicting with the data-driven model. The second row shows results estimated with mobile sensor scheduled by algorithm~\ref{alg:sparse}. The third row shows the estimation results by randomly chosen sensors. We assume the same biased guess of initial status for each case.}
    \label{fig:sparse_depth}
\end{figure*}

In Figure~\ref{fig:sparse_depth}, we qualitatively compare the estimation of floods states under different sensing strategies. The same initial guess, which is corrupted by noise from true states, is exposed to different task scheduling algorithms. We can see that the estimated flood states are closer to ground truth, in comparison to the prediction-only case. Furthermore, the informative strategy keeps a better track of the evolution of floods than choosing sensing locations randomly.

\subsection{Small number of mobile sensors together with fixed sensors}
Rivers are among the most important areas to monitor during floods~\cite{tinka2012floating}. First, most precipitation would ultimately flow into the river and cause its level to increase significantly, bringing important information about the flooding situation. Second, permanent rivers typically have water levels and surface velocities that are considerably larger than levels and surface velocities associated with temporary urban water streams during floods, yielding a higher signal to noise ratio for the flood sensors. In the second experiment, we simulate the deployment of a number of fixed sensors along the river, together with several mobile sensors. This experiment shows that algorithm~\ref{alg:sparse} makes advantage of mobile sensors to improve the estimation results beyond the fixed sensor only strategy. Meanwhile, both quantitative and qualitative analyses are given to illustrate how the prediction horizon in algorithm~\ref{alg:sparse} affects the estimation results and planned trajectories.

\begin{figure*}
        \centering
        \begin{subfigure}[b]{0.4\textwidth}
            \centering
            \includegraphics[width=\textwidth]{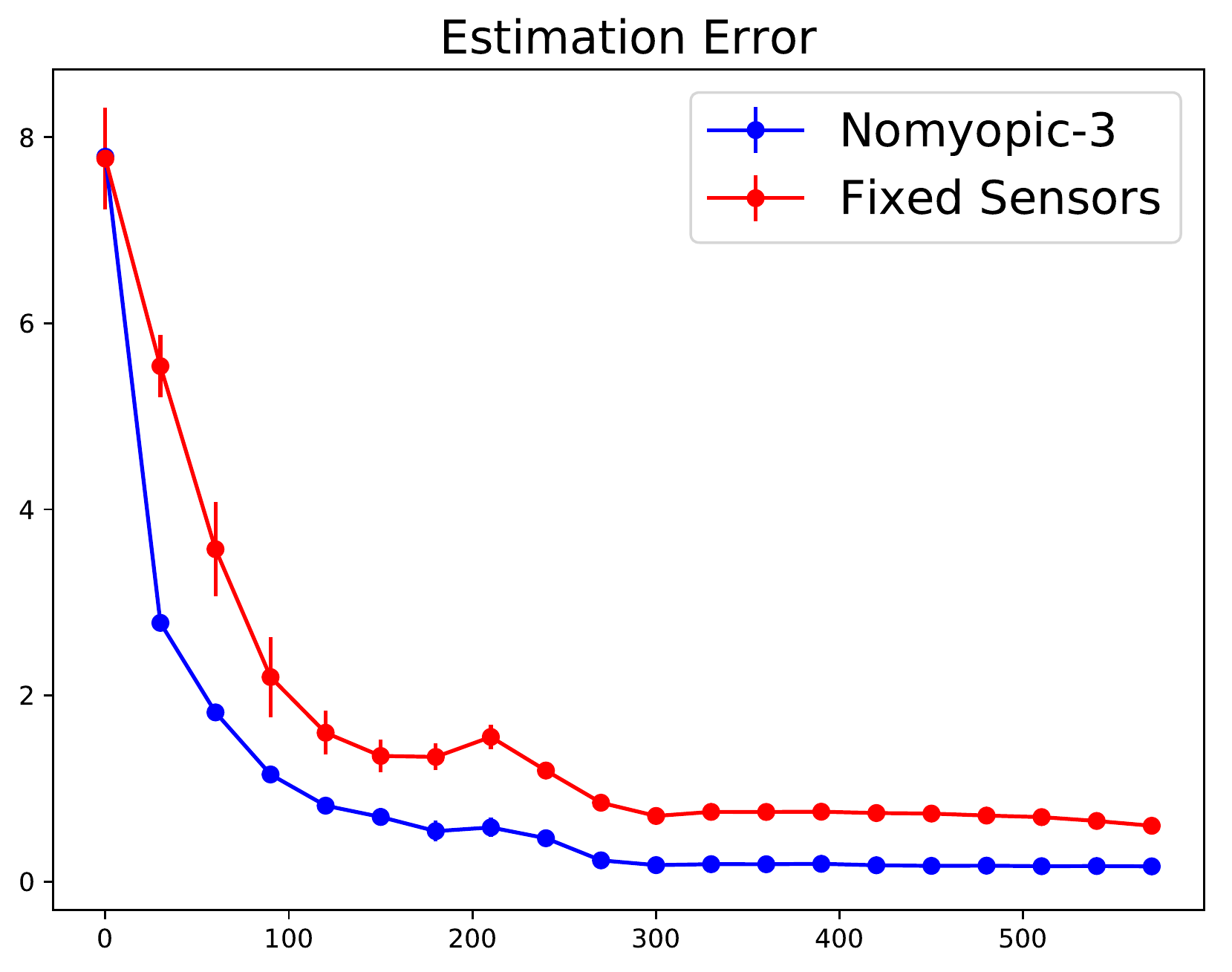}
            \caption[]%
            {{\small }}    
            \label{fig:mean and std of net14}
        \end{subfigure}
        \quad
        \begin{subfigure}[b]{0.4\textwidth}  
            \centering 
            \includegraphics[width=\textwidth]{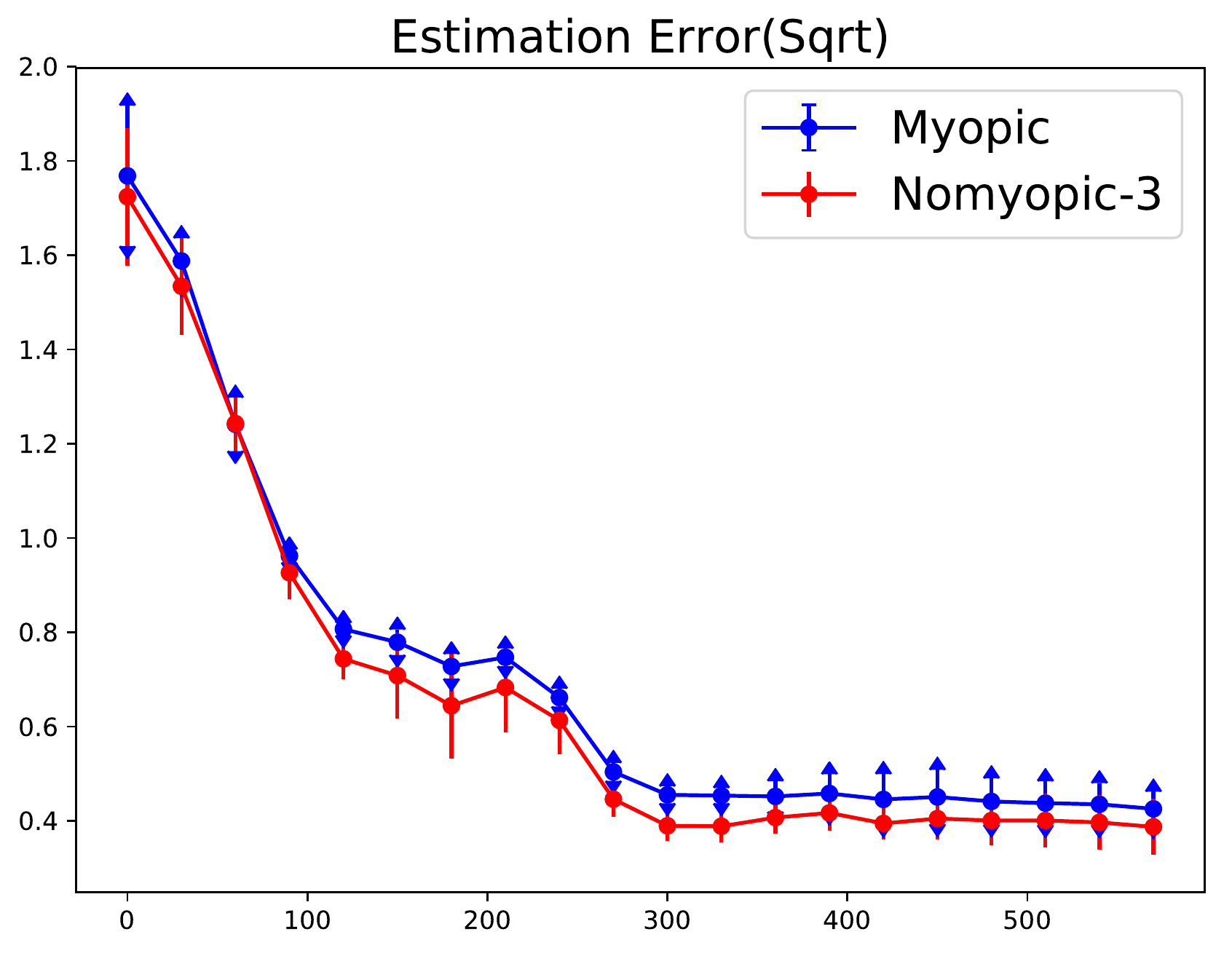}
            \caption[]%
            {{\small }}    
            \label{fig:mean and std of net24}
        \end{subfigure}
        \vskip\baselineskip
        \caption[ The Analysis of performance with a few mobile sensors and a bunch of fixed sensor. ]
        {\small Performance comparison of mobile sensing and fixed sensing. In this figure, we use the same data as in Figure~\ref{fig:single}. (a) shows the estimation error with fixed sensors only and the result by mobile sensors working together with fixed sensors. (b) shows the comparison between setting different time horizons in algorithm~\ref{alg:sparse} when scheduling mobile sensor tasks. 1 step(myopic), 3 steps(non-myopic) are compared.} 
        \label{fig:fixedmobile}
\end{figure*}

\begin{figure}
            \centering
            \includegraphics[width=0.5\textwidth]{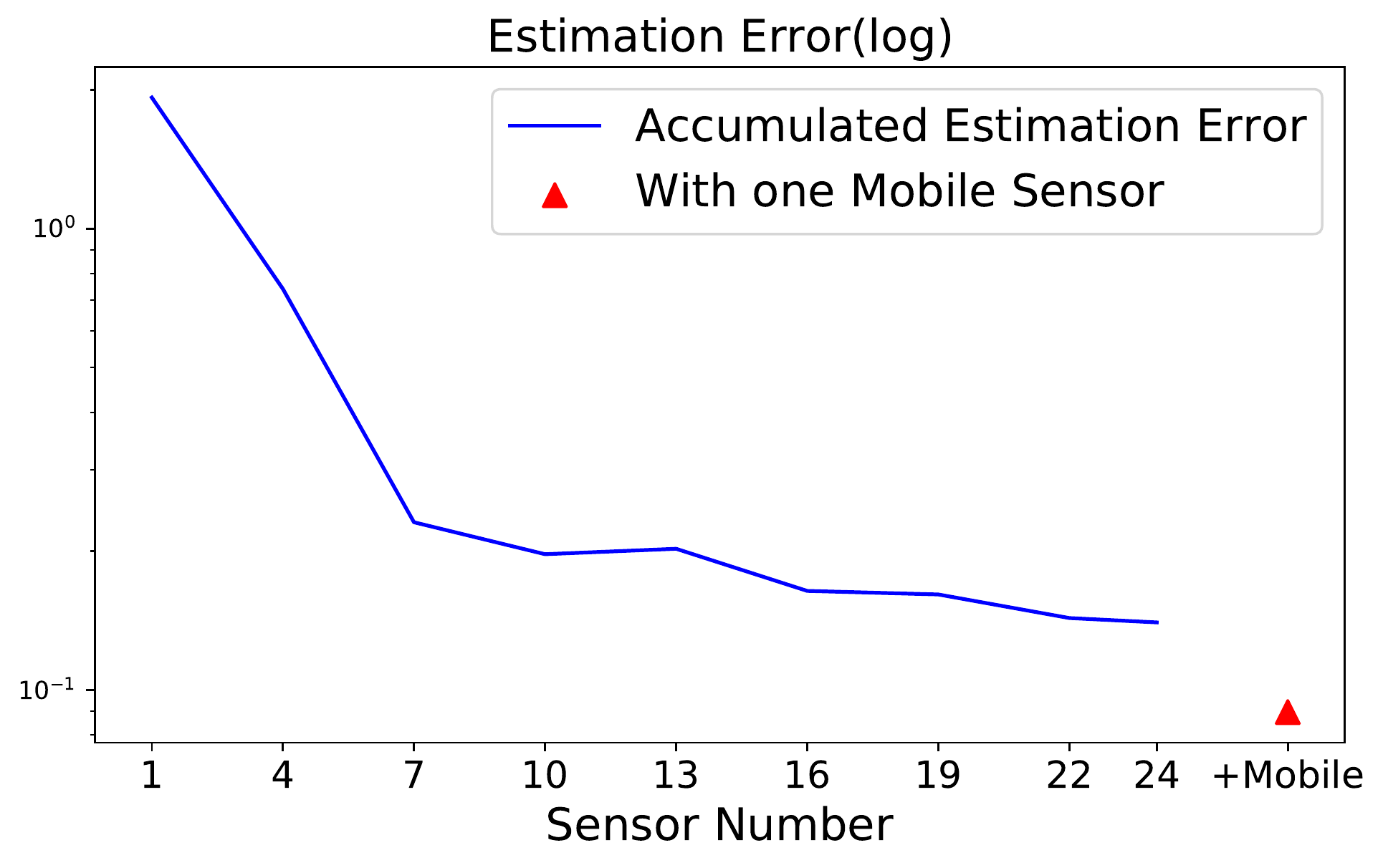}
            \caption{The comparison of estimation errors by different number of sensors. We gradually increase the number of fixed sensors and plot the estimation error changes (X-axis 1 - 24). When increasing fixed sensors cannot reduce the estimation error significantly, we add a mobile sensor to the group of fixed sensors (+Mobile). }
            \label{fig:sensor_number}
\end{figure}

In Figure~\ref{fig:fixedmobile}(a), it can be seen that, qualitatively, adding mobile sensors to an existing fixed sensor network would significantly improve the estimation performance. In order to quantify the added value of mobile sensor data (in addition to existing fixed sensor data), we compare the performance improvement using different number of fixed sensors in Figure~\ref{fig:sensor_number}. We can see that while increasing the number of fixed sensors could improve the estimation performance when the number of sensors is less than 10, while the improvement is almost non-existent when the number is greater than 20. However, we can see from the difference between last two points in Figure~\ref{fig:sensor_number} that adding one mobile sensors to 24 fixed sensors could continue to improve the estimation performance greatly. Figure~\ref{fig:fixedmobile}(b) compares the performance of algorithm~\ref{alg:sparse} with different prediction horizons. Planning under the non-myopic strategy would reduce the estimation error comparing to planning under myopic strategy.

\begin{figure}
            \centering
            \includegraphics[width=0.5\textwidth]{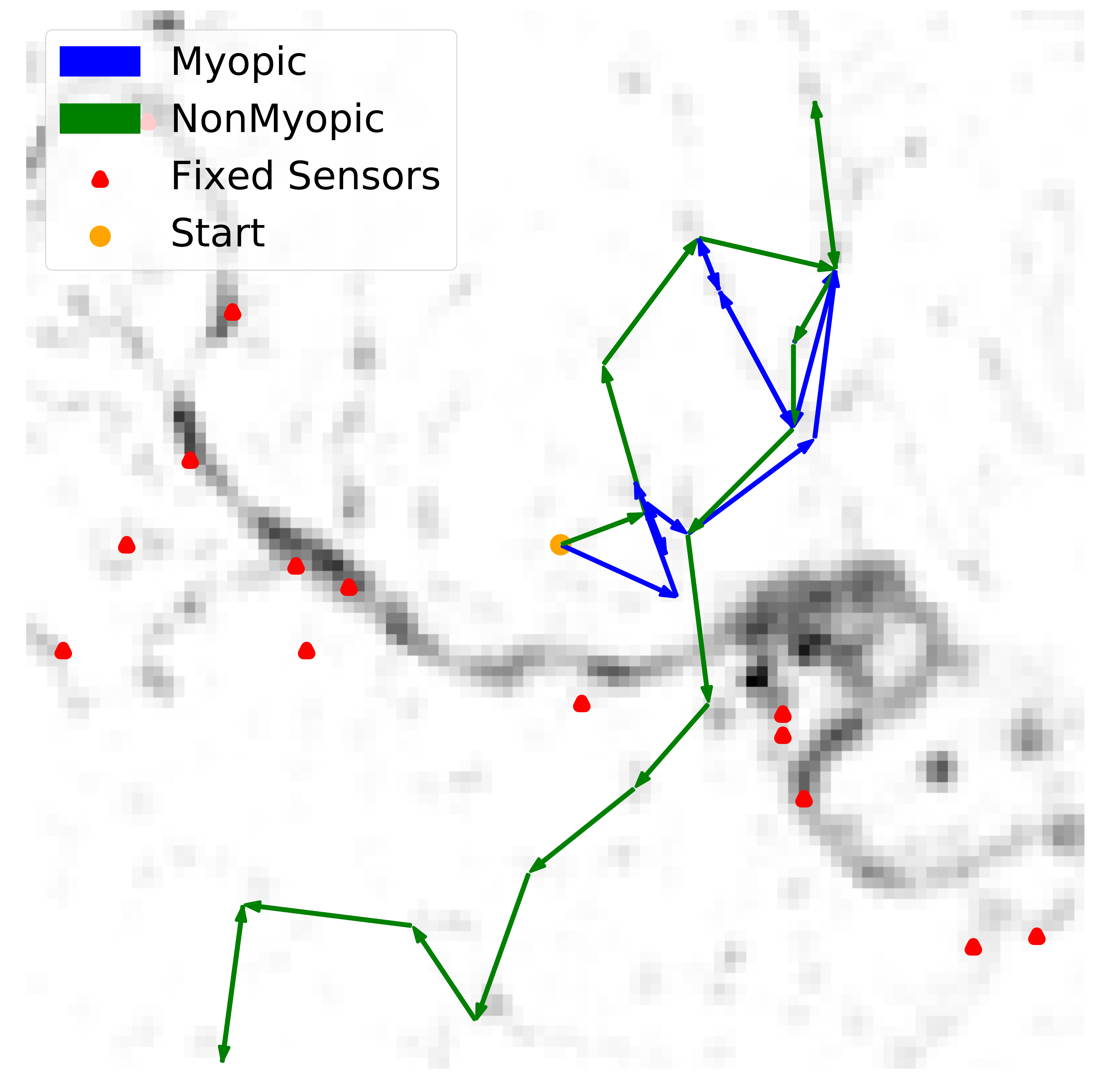}
            \caption{The comparison of mobile sensor tracks when planning under myopic and nonmyopic strategies. Red dots are the locations of fixed sensors, the blue track is induced by the myopic strategy, while the green track is induced by the nonmyopic strategy. In this Figure, all mobile sensors start from the same location, and the initial states and inputs are identical.}
            \label{fig:track}
\end{figure}

\begin{figure*}
            \centering
            \includegraphics[width=0.9\textwidth]{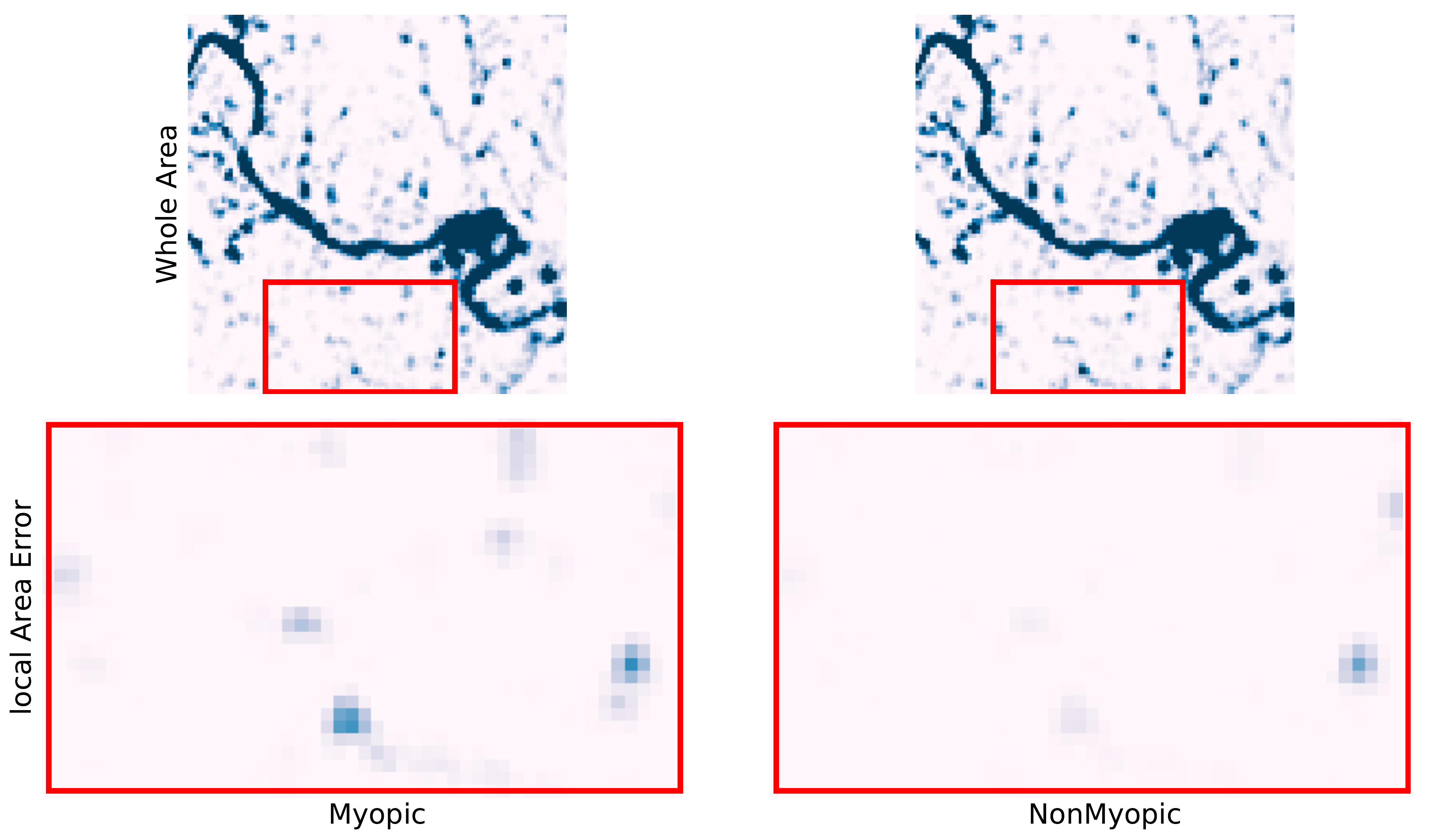}
            \caption{The demonstration of estimation results corresponding to the same experiment case in Figure~\ref{fig:track}. The first row shows the estimation result of whole domain. The second row shows the estimation error in a local area for each algorithm. The first column shows the estimation result following myopic strategy. The second column shows the estimation result following nonmyopic strategy.}
            \label{fig:fixedmobiledepth}
\end{figure*}

In order to understand qualitatively the results in Figure~\ref{fig:fixedmobile}, we plot the scheduled trajectories of mobile sensors by the algorithm with different predicting horizons. According to Figure~\ref{fig:track}, we can see that the green track which is planned under the non-myopic strategy is guided around the whole interested domain. However, the blue track which is planned under the myopic strategy is trapped in a local area. This is a reasonable result as myopic strategies can not see other areas taht possibly have higher rewards during the plan. It also indicates that the function $\Gamma(\omega)$ successfully models the causal relation between sequential observation actions and encourages sensors to explore uncorrelated states. The difference in estimation results are presented in Figure~\ref{fig:fixedmobiledepth}. The non-myopic strategy gives estimation results closer to ground truth in local areas where the myopic strategy has not brought the mobile sensors to.

\subsection{Large number of mobile sensors}
% We assume that each mobile sensor will only be in charge of a local area which can be covered by the mobile sensor in one time step. Under this setting, we only need to figure out the optimal combination of measurement locations at each time step and do not need to worry about the restriction by sensor moving. We would like to follow the greedy algorithm proposed by~\cite{krause2008near} to choose the set of measurement points at each time step. The algorithm is shown in \textcolor{red}{(algorithm psedo code)}.

% \begin{algorithm}
%   \caption{Sensor Location Selection for Dense Case}\label{alg:dense}
%   \begin{algorithmic}[3]
%   \State \textbf{Input:} $\chi$, $\chi_1$, $\dots$, $\chi_n$, $b_S$
%   \State $\Omega = \{ \}$ \Comment{The set to save the sensor deployment points.}
%   \For{$i = 1:n:$}
%     \State $s_i = \arg \max_{x_j \in \chi_i} r(b_S, \Omega + \{x_j\})$
%     \State $\Omega = \Omega + s_i$
%   \EndFor
%   \State \textbf{Return: }$\Omega$
%   \end{algorithmic}
% \end{algorithm}

In this experiment, $24$ mobile sensors are deployed. We follow the algorithm~\ref{alg:alg-multiple} to schedule tasks for mobile sensors. To compare with the performance of mobile sensors, we experiment with another $30$ different sets of fixed sensors. Each set of fixed sensors is also made up of $24$ sensors and they are chosen according to simulation data and metrics by~\cite{krause2008near, joshi2008sensor}. 

\begin{figure*}
        \centering
        \begin{subfigure}[b]{0.4\textwidth}
            \centering
            \includegraphics[width=\textwidth]{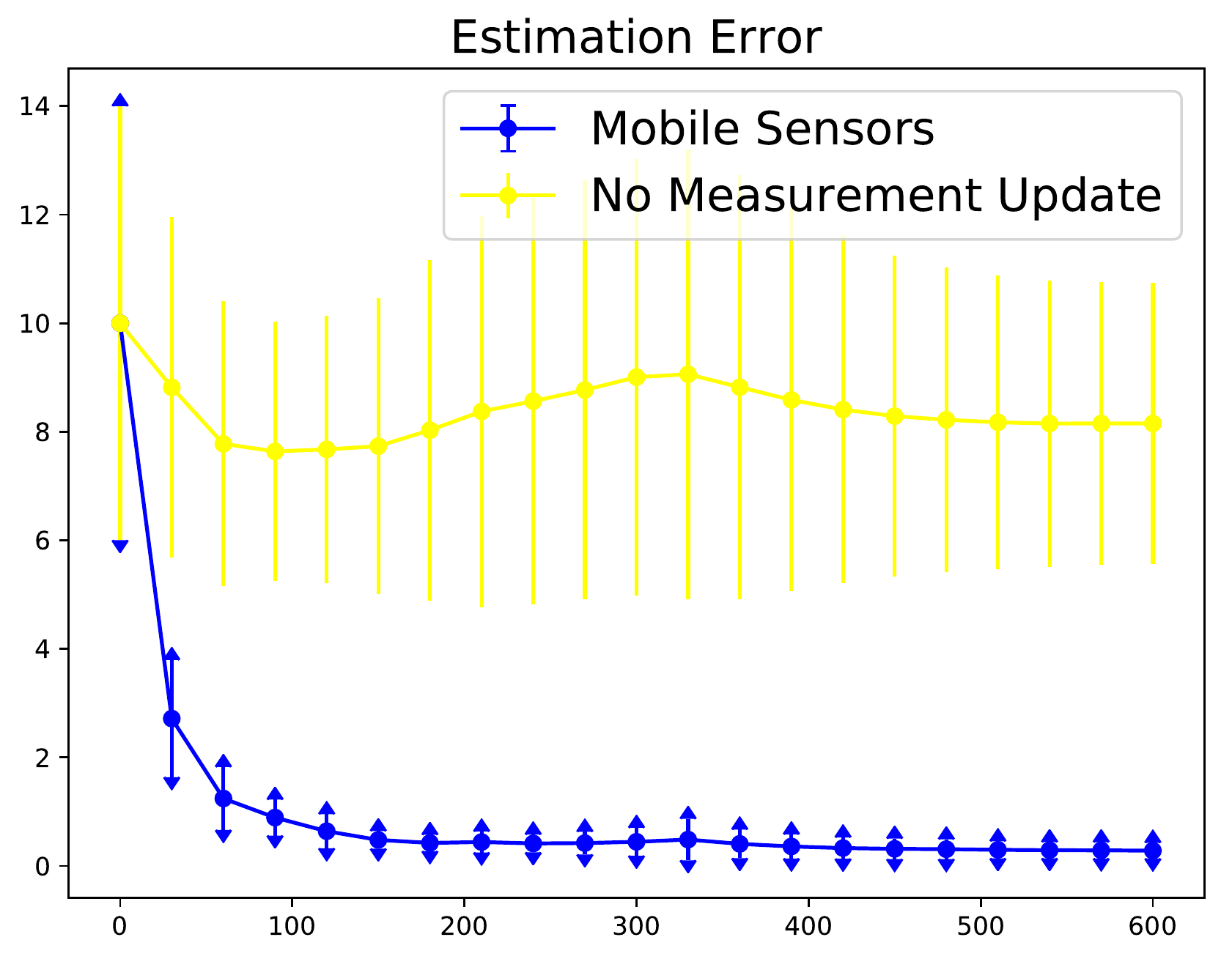}
            \caption[]%
            {{\small }}    
            \label{fig:mean and std of net14}
        \end{subfigure}
        \quad
        \begin{subfigure}[b]{0.4\textwidth}  
            \centering 
            \includegraphics[width=\textwidth]{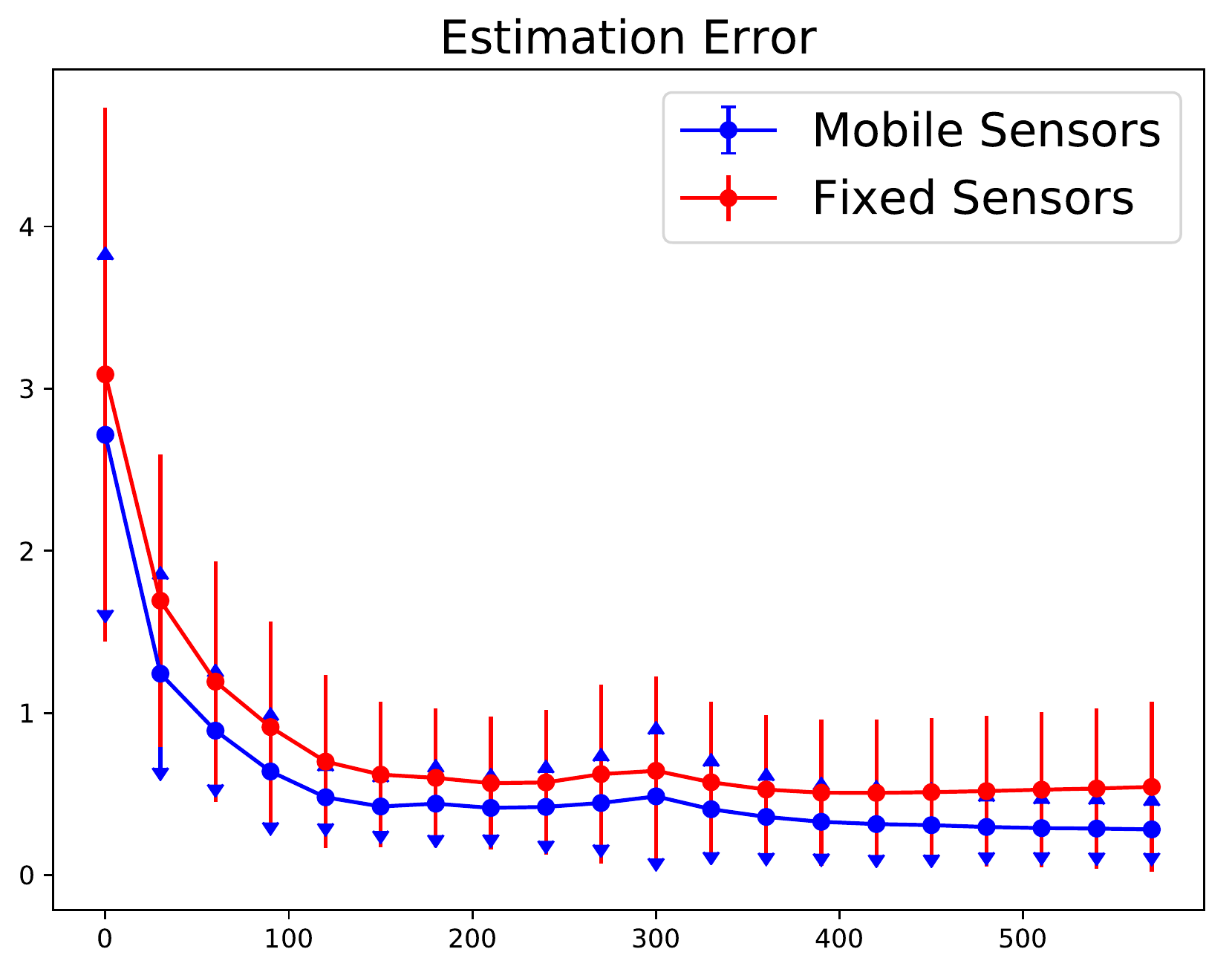}
            \caption[]%
            {{\small }}    
            \label{fig:mean and std of net24}
        \end{subfigure}
        \vskip\baselineskip
        \begin{subfigure}[b]{0.4\textwidth}   
            \centering 
            \includegraphics[width=\textwidth]{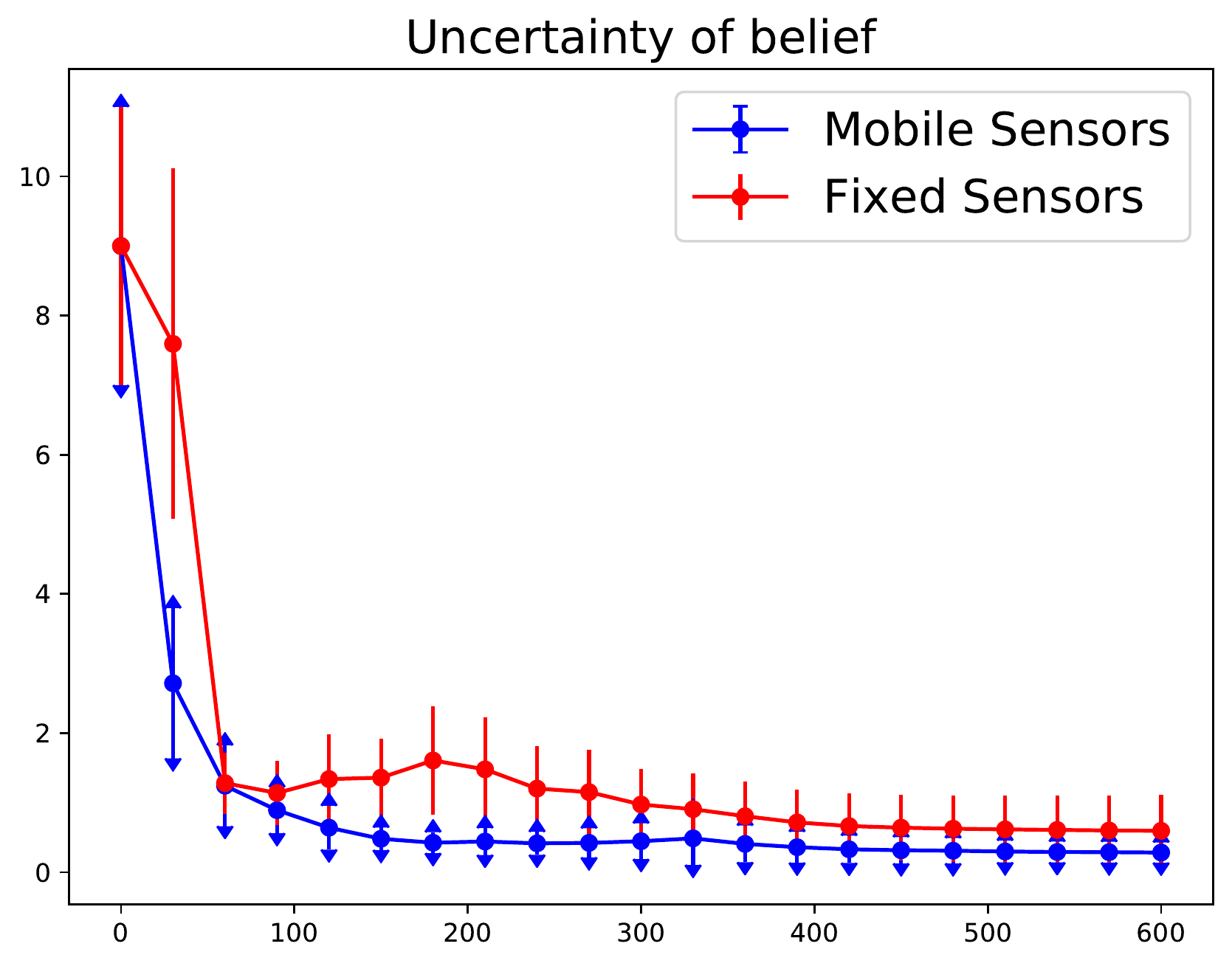}
            \caption[]%
            {{\small }}    
            \label{fig:mean and std of net34}
        \end{subfigure}
        % \quad
        % \begin{subfigure}[b]{0.3\textwidth}   
        %     \centering 
        %     \includegraphics[width=\textwidth]{figs/exp1_infogain.pdf}
        %     \caption[]%
        %     {{\small }}    
        %     \label{fig:mean and std of net44}
        % \end{subfigure}
        \caption[Quantitative Performance analysis of estimation results with large number of sensors.]
        {\small Quantitative Performance analysis of estimation results with large number of sensors. Experiments data used here is the same as the data used in Figure~\ref{fig:single}. (a) shows the comparison of estimation results between algorithm~\ref{alg:alg-multiple} and the results without any measurement updates. (b) shows the comparison of estimation error between fixed sensors and mobile sensor strategies. In order to show the comparison clearly, we plot from the second step. (c) shows the comparison in term of uncertainty of estimation.} 
        \label{fig:dense}
\end{figure*}

According to the results shown in Figure~\ref{fig:dense}, we can see that in terms of estimation error, both the fixed sensors deployment and mobile sensors strategy could improve a lot over the prediction only case. Mobile sensors do better than fixed sensors in reducing the estimation error and uncertainty of the system. Meanwhile, it is clear that even after a long time, the estimation results by fixed sensors are still not as good as estimation results by mobile sensors. The reason is that mobile sensors are keeping exploring new areas and thus keep reducing the error. Although these sets of fixed sensors are picked to maximize specific metrics, they can not actively exploring the region thus would miss information in some local areas.

\section{Conclusion}
\label{sec:conclusion}
In this paper, we propose a real-time mobile sensor task scheduling algorithm to arrange mobile sensors to monitor city-scale environmental hazards. The algorithm is based on an approximation of the forward search tree. The approximated search tree keeps the full prediction power of the dynamical model, which is data-driven, based on a Partial Differential Equation. Functions based on the correlation between observation actions are added to the approximated search tree to account for the missed causal relationship between observations due to omitting the data assimilation steps in the search tree. The approximated tree thus decouples the term $D_{obs}^{T+1}$ which grows exponentially with the prediction horizon and the time-consuming data assimilation term. Experiments show that the algorithm is real-time on regular desktop computers for city-scale high dimensional systems, and achieves near-optimal performance in reducing estimation error and uncertainty. The scheduled paths for mobile sensors can explore actively around the entire spatial domain, which overcomes the drawbacks of myopic strategies. The experiments results also indicate that a precise data-driven model could track the development of covariance of the system well such that assimilation could improve the estimation precision significantly. In future work, the authors are interested in two possible research directions. First, the authors plan to explore sensor management strategies in which the environmental dynamics are modeled in a continuous manner, as a distributed parameters system. Modeling the dynamics of the environment continuously could benefit the problem by reducing the dimension of the state, through the choice of a sparse set of functions to represent it. Second, the authors would like to explore searching for optimal or suboptimal solutions of the same sensing problem, by modeling the problem as a Markov Decision Process (MDP). The authors hope that new progress in MDP solvers or reinforcement learning could help find near-optimal results to this large scale estimation/sensor placement problem.

\section*{Acknowledgement}
The Authors would like to acknowledge NSF(National Science Foundation) CPS No.1739964 and NSF CIS No.1636154 for research Funding. 

\newpage
% \setcitestyle{square}
\bibliographystyle{elsarticle-num}
\section*{References}
\bibliography{citations}

\end{document}